\documentclass[pageno]{jpaper}

\usepackage[normalem]{ulem}

\usepackage{ifthen}
\usepackage[dotinlabels]{titletoc}
\usepackage{colortbl}
\usepackage{array}

\usepackage{microtype}

\usepackage{listings}
\usepackage[dvipsnames]{xcolor}
\usepackage{inconsolata}

\usepackage{booktabs} 

\definecolor{codeblue}{rgb}{0.2,0.2,1}
\definecolor{codegreen}{rgb}{0,0.6,0}
\definecolor{codegray}{rgb}{0.5,0.5,0.5}
\definecolor{codepurple}{rgb}{0.58,0,0.82}
\definecolor{backcolour}{rgb}{0.95,0.95,0.92}
\lstdefinestyle{mystyle}{
    commentstyle=\color{codegreen},
    keywordstyle=\color{codeblue},
    numberstyle=\tiny\color{codegray},
    stringstyle=\color{codepurple},
    basicstyle=\scriptsize\ttfamily,
    breakatwhitespace=false,         
    breaklines=true,                 
    captionpos=b,                    
    keepspaces=true,                 
    numbers=left,                    
    numbersep=5pt,                  
    showspaces=false,                
    showstringspaces=false,
    showtabs=false,
    frame=single,
    tabsize=4,
    columns=flexible
}

\hypersetup{pdfborder=0 0 0, colorlinks=true, citecolor=red, urlcolor=black, linkcolor=black}

\newcommand{\IGNORE}[1]{}

\usepackage[final]{ifdraft}

\ifdraft{
\newcommand{\COMMENT}[3]{{\color{#1}{[}{#2: #3}{]}}}
}{
\newcommand{\COMMENT}[3]{}
}

\newcommand{\JP}[1]{\COMMENT{cyan}{JP}{#1}}

\newcommand{\code}[1]{\text{\lstinline[basicstyle=\fontsize{9pt}{9.25pt}\ttfamily]{#1}}}

\title{Programming Heterogeneous Systems from an Image Processing DSL}

\author{
{\large \textrm Jing Pu, Steven Bell, Xuan Yang, Jeff Setter, Stephen Richardson}\\
{\large \textrm Jonathan Ragan-Kelley\textsuperscript{\ensuremath\dagger}, Mark Horowitz} \\
{\textrm Stanford University \textsuperscript{\ensuremath\dagger}UC Berkeley} \\
{\small \{jingpu,sebell,xuany,setter,steveri,horowitz\}@stanford.edu, \textsuperscript{\ensuremath\dagger}jrk@berkeley.edu }
}

\begin{document}
\date{}
\maketitle
\thispagestyle{empty}


\begin{abstract}

Specialized image processing accelerators are necessary
to deliver the performance and energy efficiency required by important applications in 
computer vision, computational photography, and augmented reality.
But creating, ``programming,''and integrating this hardware into a hardware/software system is difficult.
We address this problem by extending the image processing language {\em Halide} so users can specify which portions of their applications should become hardware accelerators, and then 
we provide a compiler
that uses this code to automatically create the
accelerator along with the ``glue" code needed for the user's application to access this hardware.
Starting with Halide not only provides a very high-level functional description of the hardware,
but also allows our compiler to generate the complete software program including the sequential part of the workload, which accesses the hardware for acceleration.
Our system also provides high-level semantics to explore different mappings of applications
to a heterogeneous system, with the added flexibility of being able to map at various throughput rates.

We demonstrate our approach by mapping applications to a Xilinx Zynq system.
Using its FPGA with two low-power ARM cores, our design achieves
up to 6$\times$ higher performance and 38$\times$ lower energy 
compared to the quad-core ARM CPU on an NVIDIA Tegra K1, 
and 3.5$\times$ higher performance with 12$\times$ lower energy 
compared to the K1's 192-core GPU.

\end{abstract}


\section{Introduction}
\label{sec:intro}

The performance and energy efficiency of image processing tasks are becoming increasingly
important as cameras become ubiquitous, and as our ability to extract 
information from images improves.
These tasks are extremely computationally intensive, 
requiring, for example, 120 gigaops/sec
to process 1080p/60fps raw video~\cite{hegarty2014siggraph}.
To efficiently process so many pixels, designers historically built custom hardware engines specialized to the task.
For example, a typical image signal processor (ISP) in a mobile SoC operates at 1.2 giga\-pix\-els/sec~\cite{qualcomm800}
and hardware video codecs perform an equally immense amount of processing, both with power budgets low enough for a smartphone to run for hours on a small battery.
This efficiency is possible because the applications have extreme data locality, and matching the hardware to 
their computation patterns can yield enormous energy savings.

The problem with these platforms is the brittle nature of the functions provided by the accelerators.
These functions are relatively fixed, in order to keep their performance and efficiency high, 
so their utility to an application programmer is limited to predefined library calls. 
However, application demands are rapidly evolving, and a more flexible approach is needed.
Configurable hardware, using either coarse grain reconfigurable arrays (CGRA)~\cite{mei2003adres, govindaraju2011dynamically} or FPGAs~\cite{hauck2010reconfigurable} are one approach to providing a flexible machine that can be configured to match the data flow of different algorithms. 
Although these architectures promise much better energy efficiency
compared to CPUs or GPUs,
programming and integrating them into complete real world systems
remains a formidable task for application developers.

To help address the programming part of this challenge, C-based {\em high-level synthesis} (HLS) has been widely studied in past decades~\cite{martin2009hls, HLS4FPGA, zhang2008autopilot}. 
C HLS tools raise the design level by decoupling clock timing and automatically scheduling pipelines and other resources.  However, designers still need to create a good microarchitecture in their C-code in order to develop high performance implementations, which requires hardware expertise. 

So, to further reduce the hardware knowledge a designer needs, researchers created {\em domain specific languages} (DSLs), which can embed microarchitecture knowledge for a specific application domain in the compiler. These DSL systems, including Darkroom~\cite{hegarty2014siggraph} and HIPAcc~\cite{hipaccvivado} for imaging, can then 
to generate efficient FPGA and ASIC designs from high-level image codes~\cite{auerbach2010lime, milder2012spiral, george2014hardware, prabhakar2015generating}. 


\begin{figure}
    \centering
    \includegraphics[width=.9\columnwidth]{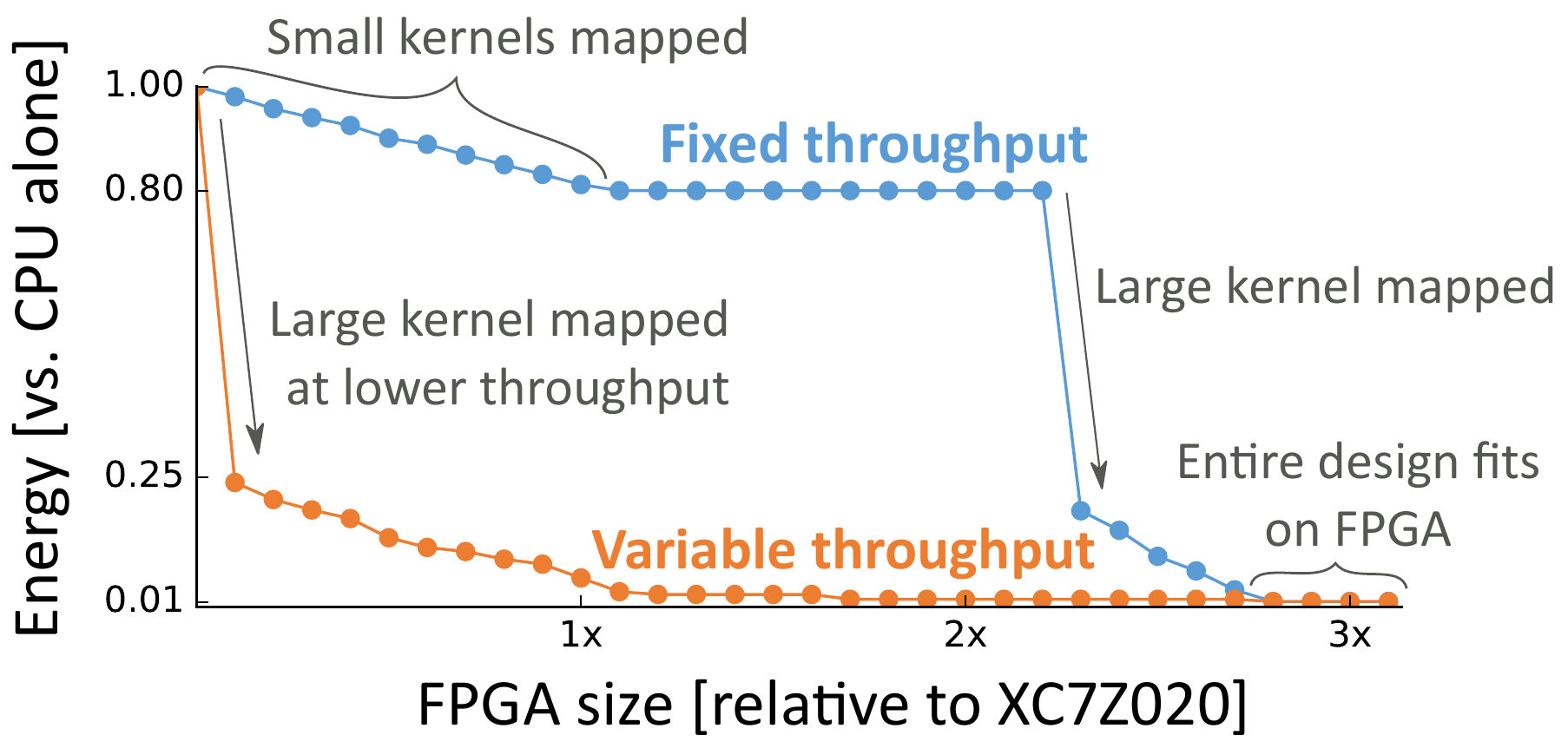}
    \caption{Energy savings from accelerating a large application composed of one large kernel computing depth from stereo plus 19 small convolution kernels on a CPU/FPGA system at different FPGA area limits.
    When accelerators run at unit rate or not at all, it takes significant FPGA resources 
    to get energy savings
    (blue).  By contrast, our variable-rate system provides benefits even from very small FPGA fabrics (orange).
    }
    \label{fig:energy-area}
\end{figure}

This paper builds on current DSL-to-FPGA systems like Darkroom and HIPAcc and extends them in three important ways. First, rather than creating a specialized language for our DSL, we use a widely popular open-source DSL, {\em Halide}~\cite{ragan2012decoupling, ragankelley2013pldi}, to describe our applications. In addition to giving us a large collection of existing image-related algorithms, it forces us to extend the microarchitecture template and compiler techniques used in previous systems to support these real-world applications.
For example, we handle the affine indices and data dependent reduction often found in kernels like downsampling or histogram.

Second, in addition to the one 
pixel/cycle 
pipelines 
of prior systems,
we can specify and generate kernels with variable throughput rates,
exploiting space-time tradeoffs often needed to map real applications on FPGAs with finite hardware. 
\autoref{fig:energy-area} dramatically illustrates the advantage of this flexibility for accelerating large applications.
At the previous one-pixel/cycle fixed throughput (blue), 
no significant energy savings is seen until the FPGA is large enough to accelerate the application's largest kernel, \emph{stereo}.  Accelerating \emph{stereo} at a much lower throughput rate requires significantly less area, so the savings are seen even with very small FPGAs (orange ``variable'' line).  As resources increase, our system gracefully tunes the throughput and includes more stages for acceleration.

Finally, our system not only generates the FPGA kernels, it also creates all the software needed to connect that hardware to the user's application. Thus in addition to the FPGA configuration file, our system creates the CPU portion of the algorithm, the Linux kernel drivers for the accelerator, and the software glue that maps user Halide calls onto kernel driver calls that access the hardware.
The automatic CPU/FPGA integration greatly helps to explore the workload partitioning between CPU and FPGA for system-level optimization, as we saw in \autoref{fig:energy-area}.

This paper makes the following contributions:
\begin{itemize}
\item 
We demonstrate that the popular image DSL {\em Halide} is sufficiently restrictive that 
much of its computation can be ``compiled'' into efficient FPGA implementations.
In fact Halide's scheduling language is powerful enough that we needed to add only two new commands to help define what and how the hardware is generated.  
This opens a large class of applications to acceleration. 

\item We extend the line buffer pipeline template of
prior image-DSL-to-hardware systems to suit the variety of computation possible in Halide. 
This includes creating pipelines of different throughputs and dealing with higher dimension input and output stencils.  In addition, our compiler implements a new loop transformation optimization, called loop perfection, to support these features.

\item We create the first end-to-end system that takes Halide user code and creates an FPGA bitstream 
%
along with a multi-threaded software program that controls the new hardware.
This end-to-end system, coupled with Halide's schedule language, allows a user to seamlessly explore the effects of moving function execution between the CPU and the FPGA.
We have used this system to implement a range of applications on a Xilinx Zynq platform.
\end{itemize}

The following section describes
tools and techniques of image processing and domain-specific languages that undergird this work. 
Sections~\ref{sec:language} and~\ref{sec:compiler} describe our extensions to the Halide language, and our compiler system that implements these extensions to produce blended CPU/FPGA designs.
Finally, we present our test platform in Section~\ref{sec:platform}, followed by an experimental comparison with other methods, and a quantitative evaluation of potential optimizations.

\IGNORE{
  In our methodology, an application is first prototyped in Halide with CPU schedules.
  Next, with a new scheduling extension described in \autoref{sec:language},
  the algorithm in the prototype can be mapped to a heterogeneous system
  (\autoref{sec:architecture}).
  A compiler tool then produces HLS C code for the specialized hardware engine,
  along with the software program running on the host (\autoref{sec:compiler}).
  The final design is implemented on a custom heterogeneous platform based on a Xilinx Zynq SoC (\autoref{sec:platform}). The next section quickly reviews some of the prior work we build upon.
}

\section{Background and Prior Work}
\label{sec:background}

Hardware designers have continually moved toward higher level language descriptions to help deal with the growing complexity of their target machines.
Since the success of early hardware description languages (HDL) such as
Verilog in the late 80's, considerable work has focused on hardware synthesis from high-level langu\-ages~\cite{martin2009hls, HLS4FPGA}.
Recent commercially-available tools include
Vivado (previously AutoPilot~\cite{zhang2008autopilot}) and Catapult~\cite{catapult}.
These tools generate hardware designs from high-level specifications in C/C++/SystemC,
decoupling the input description from issues of hardware resource allocation, clock-level timing, and pipelining.  As a result, they are seeing increasing use, especially for FPGAs. 
Yet users of these tools must still have a fairly good understanding of the microarchitecture of the hardware they want to build, since these systems can't (yet) do global restructuring of the input code to improve energy or performance.
This same limitation occurs in research efforts like Open\-CL-to-FPGA~\cite{czajkowski2012opencl},
SOpenCL~\cite{owaida2011synthesis}
and FCUDA~\cite{papakonstantinou2009fcuda}, which explore the feasibility of using GPU languages like CUDA and OpenCL as hardware description languages.

To address the limitation, we can restrict the hardware target to a single domain, like image processing, and, by examining the characteristics of applications in that domain, we can create reasonable microarchitectural templates for automatic hardware generation.

\subsection{Image Processing}
Most image processing algorithms consist of kernels operating on a small window of the image.
For example, sharpening and blurring operations can be expressed as convolutions, which use a fixed window of pixels to compute each result pixel.
Likewise, corner detection, edge enhancement, image sensor demosaicking, and color transformation all calculate their output pixels using small nearby regions of data.
In Halide, these kernels are defined as {\em functions}.
A separable $3\times3$ box blurring filter can be expressed as a chain of
two functions in $x, y$ as follows.
\begin{lstlisting}[language=C++, morekeywords={Func}, frame=none, numbers=none]
Func blury(x, y) = (input(x, y-1) + input(x, y) + input(x, y+1)) / 3;
Func blurx(x, y) = (blury(x-1, y) + blury(x, y) + blury(x+1, y)) / 3;
\end{lstlisting}
\noindent
The composition of these kernels
can be expressed as a directed acyclic graph (DAG), where the result of one computation is fed forward into the next.

This data locality, combined with the fact that image processing algorithms work on millions of pixels, means that image processing is a fertile area for code optimization and hardware acceleration.
All of the data necessary to compute a result pixel can fit within a small (and therefore near and low-power) memory block. 
Moreover, because pixels are typically computed in sequence, shared stencil data can be re-used from one pixel to the next with minimal data fetching.

However, taking full advantage of this locality while maintaining sufficient parallelism on a CPU/GPU is often difficult.
The algorithm could be computed in many ways: for example, 
%
%
the first kernel could process the whole image,
producing an entire output image to be processed by the second kernel, and so forth.
This minimizes recomputation, but has poor data locality.
Alternatively, all the kernels could be fused into a single giant computation, which is then applied pixel-by-pixel to produce the outputs directly from the inputs.
This has much better locality, but may perform many redundant computations, since every pixel is recomputed from the source, including shared values where stencils overlap.
In practice, the best performance is usually achieved with some combination of tiling (slicing the image into tiles for cache locality) and kernel fusion (computing multiple kernels before moving to the next tile), but the exact parameters are difficult to determine.



\begin{figure}
    \centering
    \includegraphics[width=.45\textwidth]{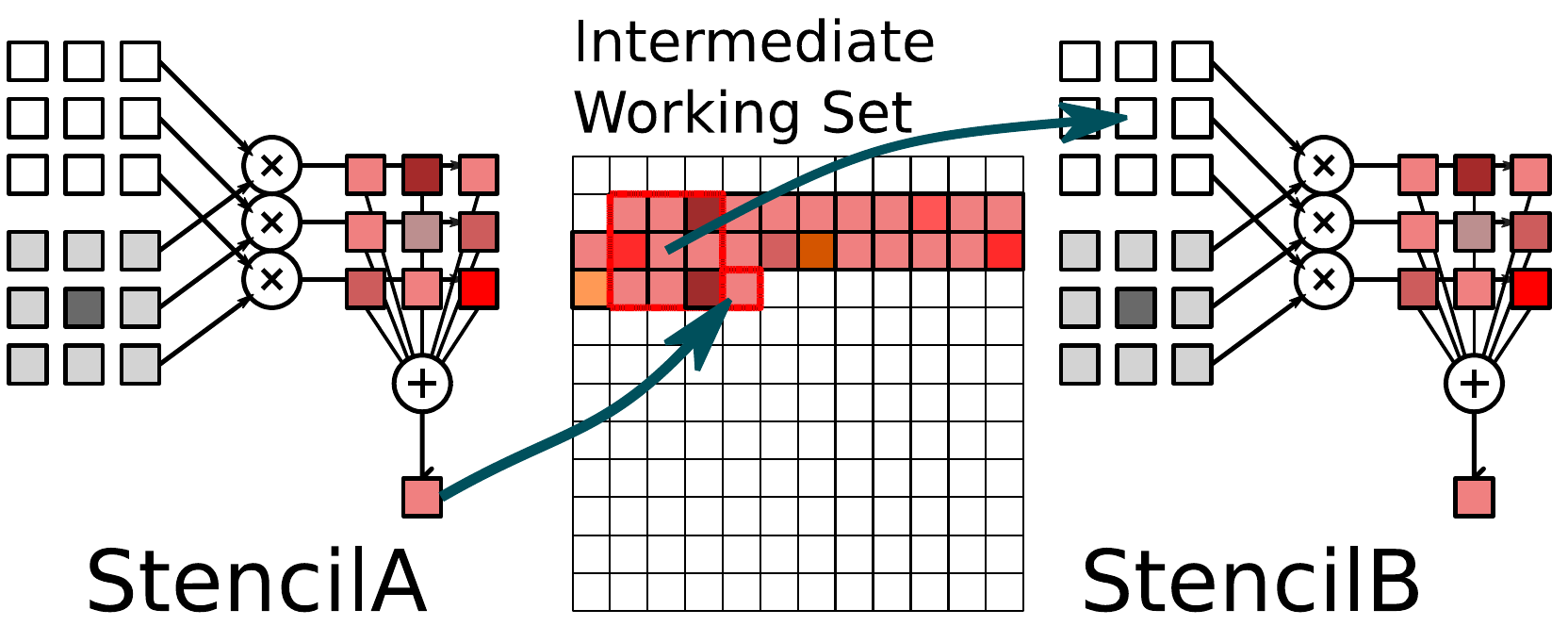}
    \caption{A line buffer captures the intermediate working set between convolution functions using minimum storage.
    }
    \label{fig:linebuffer}
\end{figure}

Custom hardware for these imaging pipe\-lines  tends to  maximize data reuse and minimize memory traffic by carefully designing pipeline stages with balanced throughput and placing specialized buffers between kernels. This organization keeps all the intermediate data in local buffers, and ensures that the image data is only fetched once at the beginning of the pipeline, and written back once at the end of the pipeline.

In \autoref{fig:linebuffer}, if the rate of new pixels produced by kernel {\em StencilA} equals the rate of $3\times3$ windows consumed by {\em StencilB,} the intermediate working set can be reduced to its minimum size of approximately two rows of pixels. A specialized buffer that captures this working set using the minimum required storage is called a \emph{line buffer}.
This line-buffered collection of deeply pipelined kernels is the microarchitectural template used by most prior image processors and image DSL-to-hardware systems, and will, with some extensions, form the basis of our design as well.

\subsection{Image DSLs}

\label{sec:related_work}

In the past decade, domain specific languages have become a popular approach to help reduce the amount of detailed knowledge required for application creation.
These systems use knowledge of a specific domain to create more efficient applications.
Spiral~\cite{milder2012spiral} synthesizes signal processing applications from
mathematical descriptions, initially for generating high-performance x86 code, and later for creating efficient hardware.
George~\cite{george2014hardware} and Prabhakar~\cite{prabhakar2015generating}
used OptiML~\cite{sujeeth2011optiml}, a machine learning DSL, to compile applications into FPGA instances.

The two prior projects most similar to our effort,
Darkroom~\cite{hegarty2014siggraph,john2015stencil} and HIPAcc~\cite{hipaccvivado}, both created image processing DSLs and provided compiler tools using a line buffered pipeline microarchitecture to guide their hardware generation. They demonstrated it was possible to take a function coded in a DSL and implement it as an FPGA or ASIC.  Recent HIPAcc followup work~\cite{ozkan2016fpga} supports vectorizing the whole pipeline.
Interestingly, HIPAcc's compiler emits HLS C and feeds that output to Vivado HLS to generate the hardware.  This flow lets them leverage the resource and datapath optimizations in the HLS tools.

Our systems builds on this earlier work.  Like HIPAcc, our compiler generates synthesizable C and leverages existing HLS tools to create the final Verilog.  To support the broad range of computation in Halide, we needed to extend the previous work in a few ways.  First, to efficiently support large programs, we needed to be able both to set the throughput rate of the hardware, and to block the image data before processing it.  
To support the wider application class available to Halide, we needed to generalize the line buffered pipeline microarchitecture to include: higher dimension input stencils ($>2$D), affine index into these arrays, and data dependent reductions. These changes required some additional analysis and optimization that is described in \autoref{sec:compiler}.


\subsection{End-to-end Systems}
Other groups have worked on creating 
complete systems to utilize hardware accelerators in a seamless way.  For example, CUDA~\cite{nickolls2008cuda} and OpenCL~\cite{stone2010opencl} provide a C-based programming environment which is similar for host and device code, along with a runtime API to manage memory transfers and synchronization.

Lime~\cite{auerbach2012lime} goes a step further by providing a unified language for CPU, GPU, and FPGA, with semantics to delineate boundaries between computation blocks.  The blocks are compiled to one or more target architectures, and then the runtime system selects a set of implementations and automatically handles data transfers and synchronization at those block boundaries.

We adopt one of the key strengths of these systems: when the compiler knows the boundary between CPU and accelerator code, it can automatically fill in the gap between them.  In our case, this is a stack of Linux executables and kernel drivers, discussed in \autoref{sec:platform}.

Our new scheduling primitives extend Lime's concept of ``relocation brackets,'' enabling the developer to easily specify whether code should run on the CPU or accelerator, separate from the algorithm itself.  However, the scheduling primitives also provide more control over the generated hardware, as described in the following sections.

\section{Language}
\label{sec:language}

\begin{figure}
    \centering
\begin{lstlisting}[language=C++, morekeywords={Func, Var}]
Func unsharp(Func in) {
  Func gray, blurx, blury, sharpen, ratio, unsharp;
  Var x, y, c, xi, yi;

  // The algorithm
  gray(x, y) = 0.3*in(0, x, y) + 0.6*in(1, x, y) + 0.1*in(2, x, y);
  blury(x, y) = (gray(x, y-1) + gray(x, y) + gray(x, y+1)) / 3;
  blurx(x, y) = (blury(x-1, y) + blury(x, y) + blury(x+1, y)) / 3;
  sharpen(x, y) = 2 * gray(x, y) - blurx(x, y);
  ratio(x, y) = sharpen(x, y) / gray(x, y);
  unsharp(c, x, y) = ratio(x, y) * input(c, x, y);
  
  // The schedule
  unsharp.tile(x, y, xi, yi, 256, 256).unroll(c)
         .accelerate({in}, xi, x)
         .parallel(y).parallel(x);
  in.fifo_depth(unsharp, 512);
  gray.linebuffer().fifo_depth(ratio, 8);
  blury.linebuffer();
  ratio.linebuffer();

  return unsharp;
}
\end{lstlisting}
    \includegraphics[width=\columnwidth]{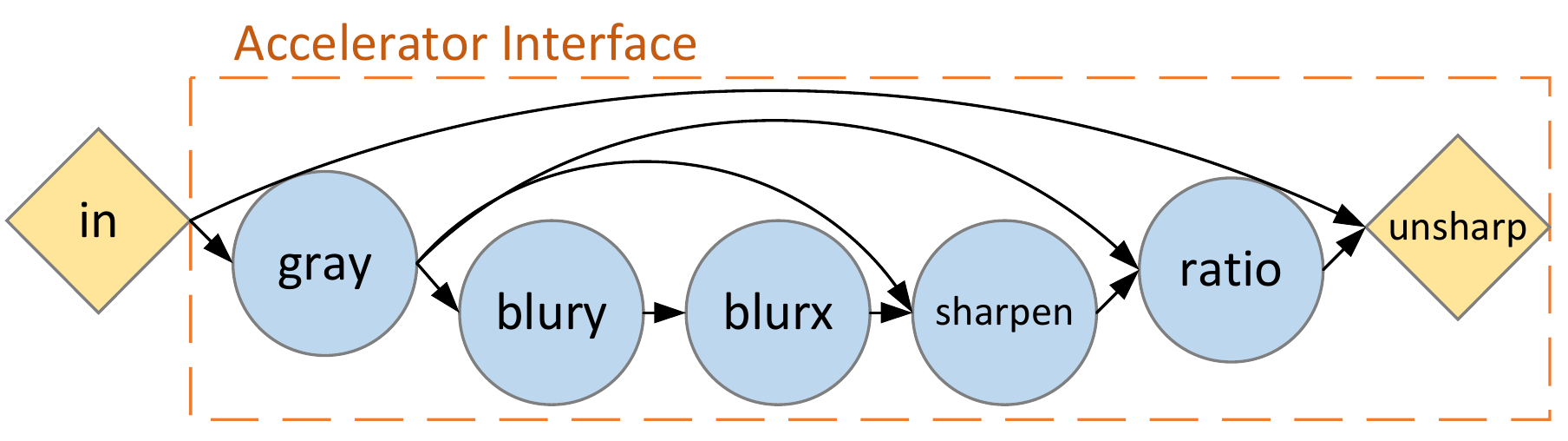}
    \caption{Algorithm and schedule code for the unsharp function, and its corresponding DAG.
    \code{accelerate} primitive defines the accelerator scope from \textit{in} to \textit{unsharp}.
    }
    \label{fig:unsharp-code}
    \vspace{-0pt}
\end{figure}

The Halide language tackles the problem of finding the most efficient implementation for an application by separating the computation to be performed (the \emph{algorithm}) from the order in which it is done (the \emph{schedule}).  The language provides \emph{scheduling primitives} which control high-level scheduling decisions like tiling, loop reordering, and parallel execution, making it easy to experiment with various tradeoffs between locality, parallelism and redundant re-computation.

Our task is to extend these
semantics to cover heterogeneous systems, mostly involved with mapping Halide functions onto a specialized hardware engine.
Specifically, the schedule should include:

\begin{itemize}
    \item The scope and interface of the hardware accelerator pipeline.
    \item The granularity of the accelerator launch task, i.e. the size of output image block
        the hardware 
        produces
        per launch.
    \item The amount of parallelism implemented in the hardware da\-ta\-path, which affects the throughput of each pipeline stage.
    \item The allocation of buffers, specifically line buffers,
        that optimally trades storage resources for less re-computation.
    \item The number of delay register slices needed to match varying computation latencies.
\end{itemize}

Many hardware scheduling choices have analogues in CPU scheduling, and Halide already has primitives to describe them.
For example, both CPU and hardware schedules must describe computation order and memory allocation.  In such cases, we reuse as many of the existing primitives as possible.
Ultimately, we were able to achieve efficient hardware mapping and hybrid CPU/accelerator execution using only two new primitives and a bit of syntactic sugar.

The language of scheduling is best explained in the context of an example.
\autoref{fig:unsharp-code} shows a simple unsharp mask filter implemented in Halide.
Unsharp masking is an image sharpening technique often used in digital image processing.
We will use this as a running example throughout the paper, as it demonstrates many important features of our system.
The code first computes a blurred gray-scale version of
the input image using a chain of three functions (\code{gray}, \code{blury}, and \code{blurx}),
and then amplifies the input based on the difference between the original image and
the blurred image.

The hardware schedule begins on line 14.
\code{unsharp.tile} 
is a standard Halide operation, which breaks an ordinary row-major traversal (defined by the \code{Var}s \code{x} and \code{y}) into a blocked computation over tiles (here, 256$\times$256 pixels).  The variables \code{xi} and \code{yi} represent the inner loops of the blocked computation which work pixel by pixel, while \code{x} and \code{y} then become the outer loops for iterating over blocks. 

With the image now broken into constant-sized pieces, we can apply hardware acceleration.
Our first new primitive is \code{f.accelerate(inputs, innerVar, blockVar)},
which defines both the scope and the interface of the accelerator and the granularity of the accelerator task.
The first argument, \code{inputs}, specifies a list of \code{Func}s for which data will be streamed in.
The accelerator will use these inputs to
compute all intermediate \code{Func}s to produce the result \code{f}.
In this example, this is the sequence of computation through \code{gray}, \code{blury}, \code{blurx}, \code{sharpen}, and \code{ratio} that produces \code{unsharp} from \code{in} (\autoref{fig:unsharp-code}, bottom).

The block loop variable \code{blockVar} defines the granularity of the computation:
the hardware will compute an entire tile of the size that \code{blockVar} counts; in this case, 256$\times$256 pixels.
The inner loop variable \code{innerVar} controls the throughput: \code{innerVar} will increment each cycle,
in this case producing one pixel each time.
To create higher-throughput hardware, we could use Halide's \code{split} primitive to split the \code{innerVar} loop
into two, and accelerate with the outer one as the hardware stride size.

Our second new primitive is \code{src.fifo\_depth(dest, n)}.  It specifies
a FIFO buffer with a depth of \code{n}, instantiated between function \code{src} and function \code{dest}.
In the unsharp example, both \code{ratio} and \code{unsharp} consume multiple data streams (\code{sharpen} is fused into \code{ratio}),
so the latency needs to be balanced across the inputs.
The optimal FIFO depths in the DAG can be solved automatically as an integer linear programming (ILP) problem~\cite{hegarty2014siggraph},
so we can eventually automate this decision,
but for now we specify and tune it by hand.

\clearpage
\code{f.linebuffer()}, our syntactic sugar for a combination of existing Halide primitives, is designed to instantiate a line buffer for function \code{f}.\footnotemark~~Without
these primitives, functions would be fused directly into other downstream functions, potentially causing re-computation when their values were reused.

\footnotetext{
Halide's native analog of the line buffer,
the {\em sliding window} pattern, 
is achieved by specifying different compute and storage levels with
\texttt{compute\_at} and \texttt{store\_at} primitives, 
and letting the compiler apply a {\em storage folding} optimization.
We overload the semantics of this same pattern if they are used in the accelerated portions.
As \texttt{accelerate} already defines the compute and storage levels,
we add the \texttt{linebuffer} sugar which needs no additional arguments.
}

The existing Halide primitive \code{f.unroll(var, factor)} is useful for optimizing hardware.
In a CPU schedule, \code{unroll}
is used to eliminate short loops and to enable optimizations on cross-iteration sharing of data.
However, in terms of hardware, since the HLS tool schedules resource for one loop iteration,
having a larger loop body through unrolling also increases the parallelism of the datapath.
It effectively duplicates the compute units in the pipeline, potentially scaling up the throughput.
In the example, \code{unroll} on $unsharp$ causes three multipliers to be instantiated
for computing three color channels simultaneously, which scales the throughput of the pipeline
from 1/3 pixel/cycle to one pixel/cycle.

All other existing Halide primitives (e.g., \code{tile}, \code{vectorize}, \code{parallel})
remain unchanged for the portion of the program mapped to software,
where Halide already provides state-of-the-art performance on ARM and x86 CPUs.
In our example, the \code{parallel} primitives on line 16 schedule multiple tile processing tasks concurrently onto multiple CPU cores.
Further details about parallel execution are discussed in \autoref{sec:platform}.

\IGNORE{
The optimized schedule of this application for an x86 CPU 
first tiles the output image into $256\times32$ chunks.
Each row of chunks is executed in parallel.
Functions $gray$, $blur\_y$ and $ratio$, which are called more than once by their
consumer functions, are precomputed at the chunk level 
to avoid re-computation,
while other functions are implicitly inlined by the default scheduling behavior.
For all functions, the computations are packed into 8-wide vectors along the $x$ dimension.
Moreover, the channel ($c$) dimension of $unsharp$ is unrolled to eliminate
the short loop.
}

\section{Compiler Implementation}
\label{sec:compiler}

\begin{figure}
    \centering
    \includegraphics[width=.8\columnwidth]{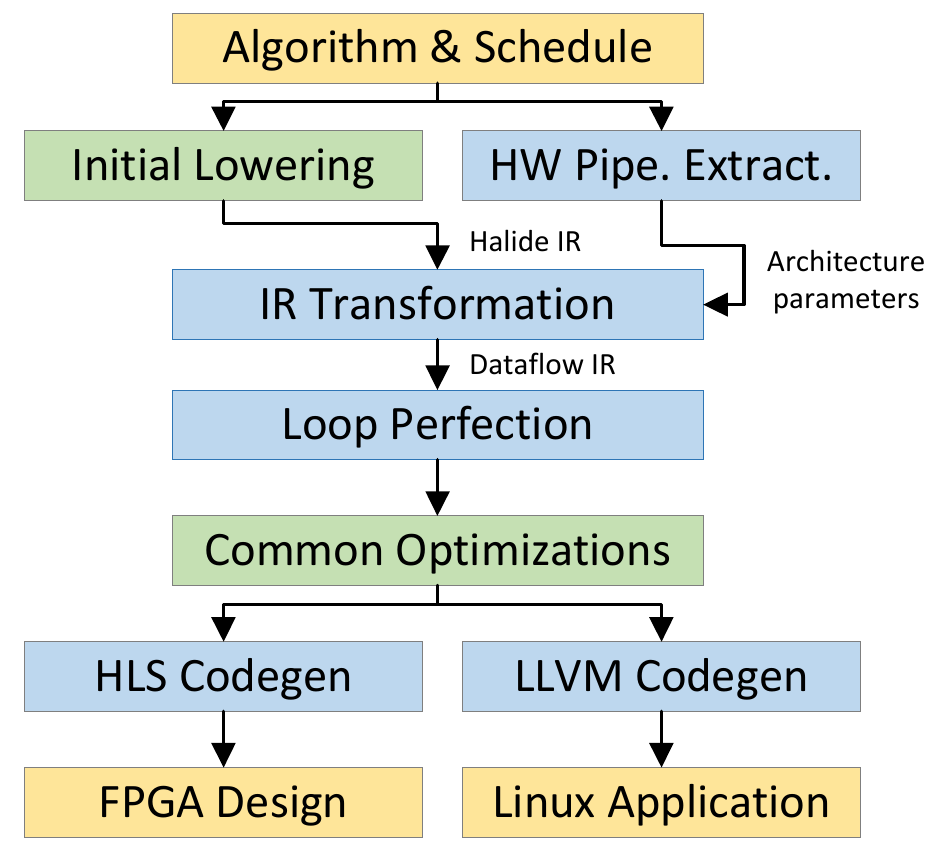}
    \caption{Compilation flow. Blue blocks are new, green blocks are unchanged/existing Halide compilation passes.
    }
    \label{fig:compiler}
\end{figure}

\autoref{fig:compiler} describes our compiler design.
The inputs to our system are an application's {\it algorithms} and {\it schedules} written in Halide.
An analysis pass extracts parameters for the architecture template.
A transformation pass re-writes the hardware parts of the Halide IR in a dataflow style.
After the loop perfection optimization and some common scalar optimization passes, like constant propagation, common sub-expression elimination etc., 
the final IR is passed to an HLS code generator and an LLVM code generator,
which produce the hardware designs and the software programs, respectively.

\subsection{Architecture Parameter Extraction}
\label{sec:extract}

\begin{figure}
    \centering
    \includegraphics[width=1.0\columnwidth]{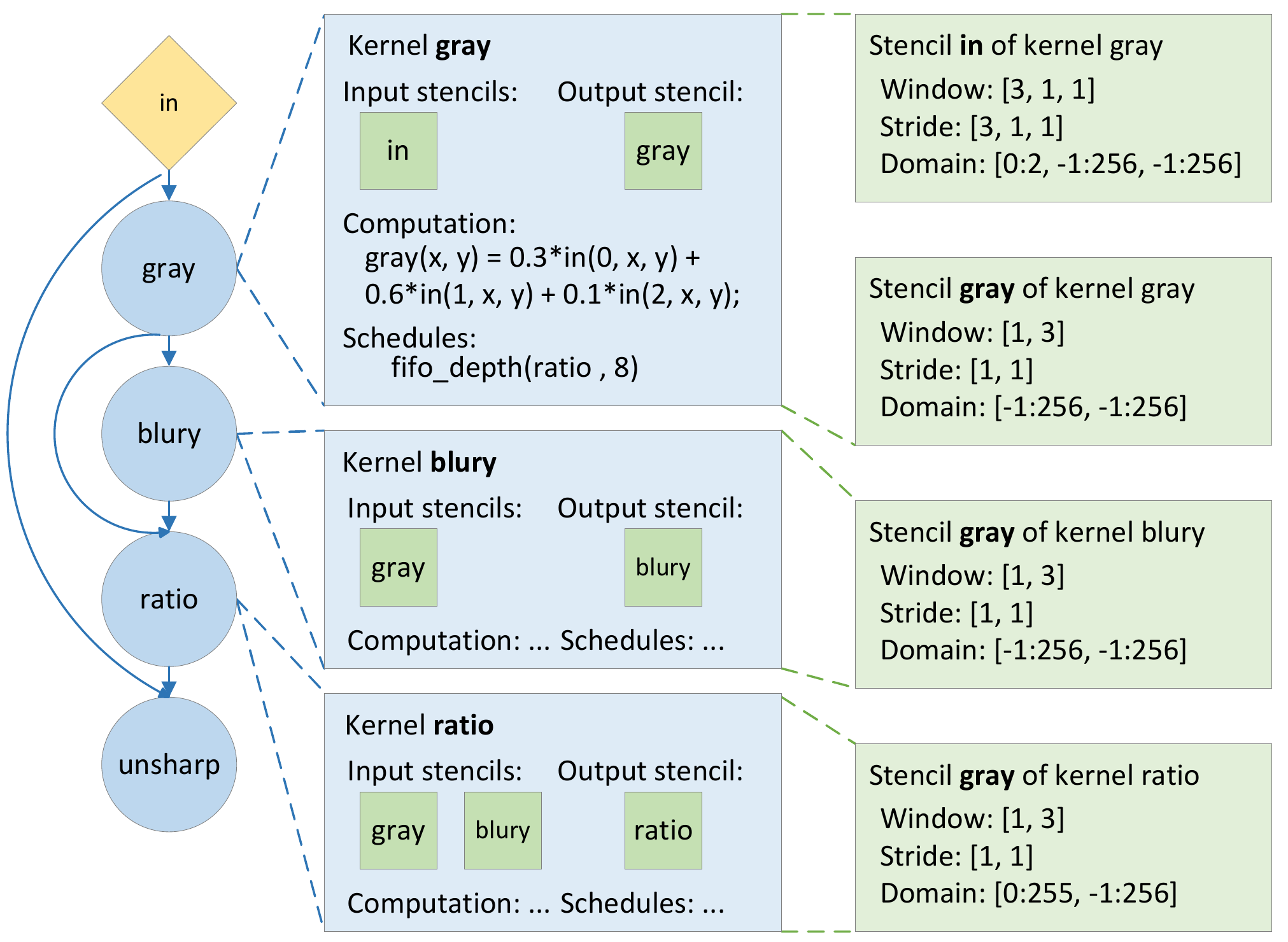}
    \caption{
      Architecture template parameters for \textit{unsharp}.
      The blue circles are stencil kernels corresponding to line buffered functions
      (\textit{blurx} and \textit{sharpen} were fused into \textit{ratio}).
      Middle and right columns show parameters of selected kernels and stencil streams.
      The domain of the output stencil stream \textit{gray} is the union of the domains of the stencil \textit{gray} in kernels \textit{blury} and \textit{ratio}.
    }
    \label{fig:unsharp-arch}
    \vspace{-0pt}
\end{figure}

Our system generates specialized hardware accelerators by instantiating architectural templates from a scheduled program.
The architectural template approximates the \textit{line-buffered pipeline} of Darkroom~\cite{hegarty2014siggraph},
with extensions to support a wider range of algorithm and performance targets.

In this template, an accelerator is a DAG whose edges are streams of windows of pixels, or {\em stencil streams},
and whose nodes are {\em stencil kernels}.
Each kernel is a Halide function scheduled for a line buffer,
into which one or more non-line-buffered functions can be fused.
%
A stencil stream is parameterized by the window size, the sliding stride, and the range of the image domain. 
Note that because an output stencil may serve as input to more than one downstream kernel, the producer kernel must compute a union of the stencils required by all consumers.
\autoref{fig:unsharp-arch} shows some of the architectural parameters for the \textit{unsharp} application of \autoref{fig:unsharp-code}.

Since the scope of the pipeline and the line buffered functions are defined in a high-level language,
composing the DAG is straightforward.
To extract the parameters of each stencil stream in the template, 
we apply bounds inference analysis recursively back from the output.
This is similar to Halide~\cite{ragankelley2013pldi}, except that we now have line buffers between stencil kernels that capture data reuse,
so the output stencil size of upstream kernels doesn't cumulatively increase.
We also apply aggressive constant propagation at this stage
in order to derive any constant bounds as early as possible.

\subsection{IR Transformation}
\label{sec:ir-transform}

\begin{figure}
    \subfloat[Halide IR of function $gray$]{
\begin{lstlisting}[language=C, morekeywords={alloc}]
alloc gray [258, 258]
for (unsharp.y, 0, 256)
  for (unsharp.x, 0, 256)
    // compute function "in" here...
    for (y, gray.y.loop_min, gray.y.loop_extent)
      for (x, gray.x.loop_min, gray.x.loop_extent)
        gray(x,y) = 0.3*in(0,x,y) + 0.6*in(1,x,y) + 0.1*in(2,x,y)
    // compute function "blury" here...
\end{lstlisting}
        \label{fig:halide-ir}
    }
    \vspace{-10pt}
    \newline
    \subfloat[Dataflow IR of kernel $gray$]{
\begin{lstlisting}[language=C, morekeywords={alloc, def_stream}]
// kernel "in" here...
def_stream gray_stream [1, 3]
def_stream gray_step_stream [1, 1] 
for (scan_y, 0, 258)
  for (scan_x, 0, 258)
    alloc in [3, 1, 1]
    alloc gray [1, 1]
    pop (in, in_stream)
    for (y, 0, 1)
      for (x, 0, 1)
        gray(x,y) = 0.3*in(0,x,y) + 0.6*in(1,x,y) + 0.1*in(2,x,y)
    push (gray, gray_step_stream)
linebuffer(gray_step_stream, gray_stream, [258, 258])
dispatch(gray_stream, [-1:256, -1:256],
         "blury", 1, [-1:256, -1:256],
         "ratio", 8, [0:255, -1:256])
// kernel "blury" here...
\end{lstlisting}
        \label{fig:dataflow-ir}
    }
    \caption{IR transformation allocates additional storage for local stencils, and separates pipeline stages with data streams.}
    \label{fig:dataflow-transform}
\end{figure}

\begin{figure}
    \centering
    \includegraphics[width=1.0\columnwidth]{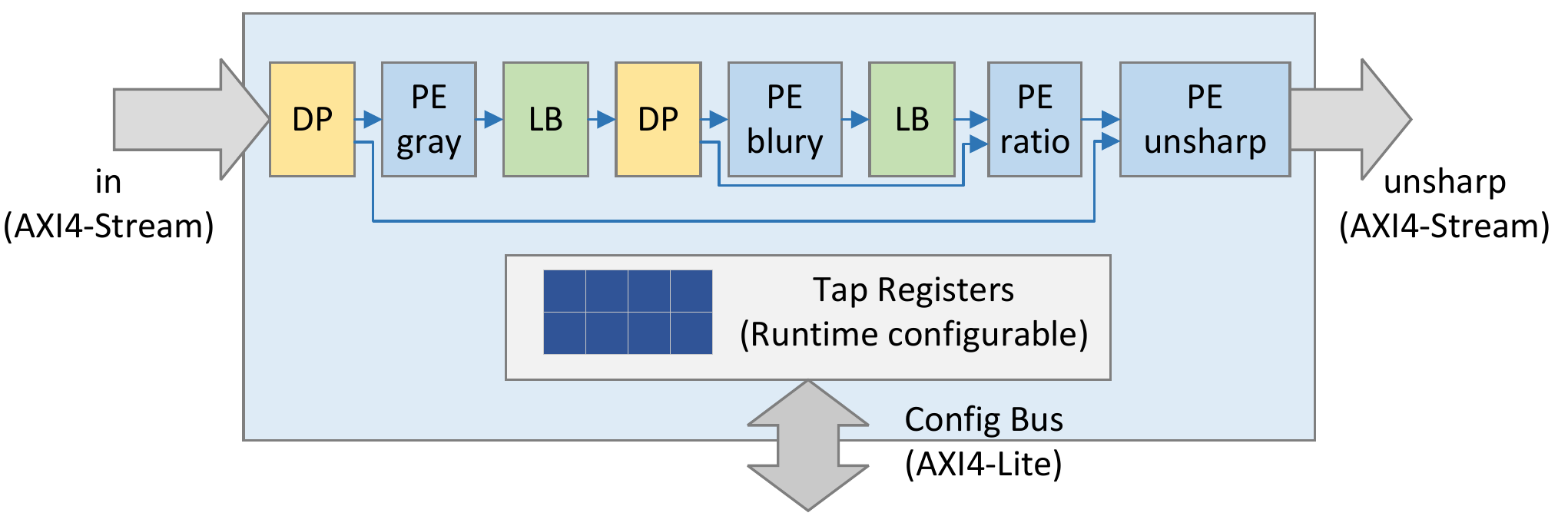}
    \caption{Pipeline implementation for \textit{unsharp}.
    Processing elements \emph{PE} implement kernel DAGS.
    Line buffers \emph{LB} capture the data reuse in the stencil pattern.
    Stream dispatchers \emph{DP} fork a stencil stream to multiple consumers.
    }
    \label{fig:unsharp-impl}
    \vspace{-0pt}
\end{figure}

Given a schedule, the Halide compiler generates an arch\-itecture-independent 
imperative representation of the algorithm,
with loop nests and storage allocations injected for each function.
\autoref{fig:halide-ir} shows part of the
Halide loop IR for function $gray$.
Because the schedule specified a sliding window order for computing $gray$,
the storage for $gray$ is allocated outside the loop nest that iterates over the 256$\times$256 block of $unsharp$,
while the computation of $gray$, along with other functions, is interleaved inside the loop nest.

To map this scheduled loop IR into the line buffer pipeline shown in \autoref{fig:unsharp-arch},
we further lower it into our own dataflow IR, using the architecture template parameters previously extracted.
First, additional storage for input and output stencils are
allocated locally to a kernel computation, and references to the original storage are replaced with references to stencils.
Then, the original storage allocation is replaced with declarations of stencil streams,
and stream operations (\code{pop} and \code{push}) are inserted before and after
the kernel computation.
Finally, loops iterating over the image domain are inserted for each compute stage,
along with calls to \code{linebuffer} and \code{dispatch} IR primitives if a line buffer or a stream dispatcher is needed.

\IGNORE{
  First, there is only one buffer allocated for each function,
  with all the producers and consumers accessing the same buffer.
  This fits a multiprocessor architecture with a cache managing the producer-consumer communication,
  but in our accelerators, each compute stage (PE) has its own local buffers, decoupled from the others, and communicates through address-free stream channels.

  Second, after line buffers are inserted between stages, compute stages are no longer interleaved in the same loop nest (over the output image), but instead each stage scans over its own image domain.
  
  To correct these issues, we transform the IR using the architecture parameters we extracted earlier.
}

\autoref{fig:dataflow-ir} shows the IR of kernel $gray$ after the transformation.
For each scanning step (here, each $scan\_x$ loop iteration), the inner loop nest computes a
$1\times1$ $gray$ stencil and pushes it to $gray\_step\_stream$, which is later line-buffered and dispatched to the kernels $blury$ and $ratio$.
The unit length loops $x$ and $y$ will be eliminated in a later simplification pass.

After the IR transformation, the dataflow IR presents a bit-accurate representation of the pipeline,
with different stages explicitly separated by data streams.
For example, the \textit{unsharp} pipeline is composed of four kernel stages, two line buffers,
and two stream dispatchers, as shown in \autoref{fig:unsharp-impl}.
Many HLS tools can infer coarse grain pipelined designs from such code structure,
including Vivado HLS with its \code{DATAFLOW} directive.
However, the throughput of stages producing less than 1\,pixel/cycle is not optimal, due to limits on the automatic loop pipelining of the current HLS tool.
We address this problem next.

\subsection{Loop Perfection Optimization}
\label{sec:loop_perfection}

\begin{figure}
    \centering
    \vspace{-6pt}
    \subfloat[Original IR.]{
      \begin{minipage}[c][78pt][b]{.42\columnwidth} 
\begin{lstlisting}[language=C, morekeywords={}, linewidth=\columnwidth]
for (scan_x, 0, 256)
    pop (in, in_stream)
    out(0) = 0
    for (d, 0, 5)
      out(0) += mask(d)*in(d)
    push (out, out_stream)
\end{lstlisting}
        \label{fig:sequential-loop-ir}
      \end{minipage}
    }
    \hspace{15pt}
    \subfloat[IR after the optimization.]{
      \begin{minipage}[c][78pt][b]{.435\columnwidth}
\begin{lstlisting}[language=C, morekeywords={}, linewidth=\columnwidth]
for (scan_x, 0, 256)
    for (d, 0, 5)
      if (d == 0) 
        pop (in, in_stream)
        out(0) = 0
      out(0) += mask(d)*in(d)
      if (d == 4) 
        push (out, out_stream)
\end{lstlisting}
        \label{fig:sequential-loop-opt-ir}
      \end{minipage}
    }
    \caption{
      Loop perfection optimization creates a larger perfect loop nest by pushing operations from the outer loop body into the innermost loop. 
    }
    \label{fig:loop_perfection}
\end{figure}

\autoref{fig:sequential-loop-ir} shows the IR of a 5-point convolution kernel scheduled at the rate of 1/5 pixel/cycle.
Loop pipelining in Vivado HLS only applies to a {\em perfect loop nest} such that only the innermost loop has operations.
Here, loops \code{scan_x} and \code{d} do not form a perfect loop nest due to the instructions on line 2, 3 and 6,
and thus loop \code{d} and line 2, 3, and 6 will run sequentially in the resulting design.
In order to generate a fully pipelined convolution stage with 1/5 pixel/cycle throughput,
the content in loop \code{scan_x} must be pushed into the innermost loop, as shown in \autoref{fig:sequential-loop-opt-ir}.

We apply restructuring automatically in the accelerated region of IR through a recursive descent algorithm.
The impact of the optimization is analyzed in \autoref{sec:loop_perfection_result}.
Note that the loop perfection is an inverse operation of loop peeling~\cite{Wolfe:1992:BIV:143095.143131},
which moves predicates out of the innermost loop to improve the performance on traditional processors.

\subsection{Code Generation}
\label{sec:codegen}

\begin{figure}
    \centering
    \includegraphics[width=0.8\columnwidth]{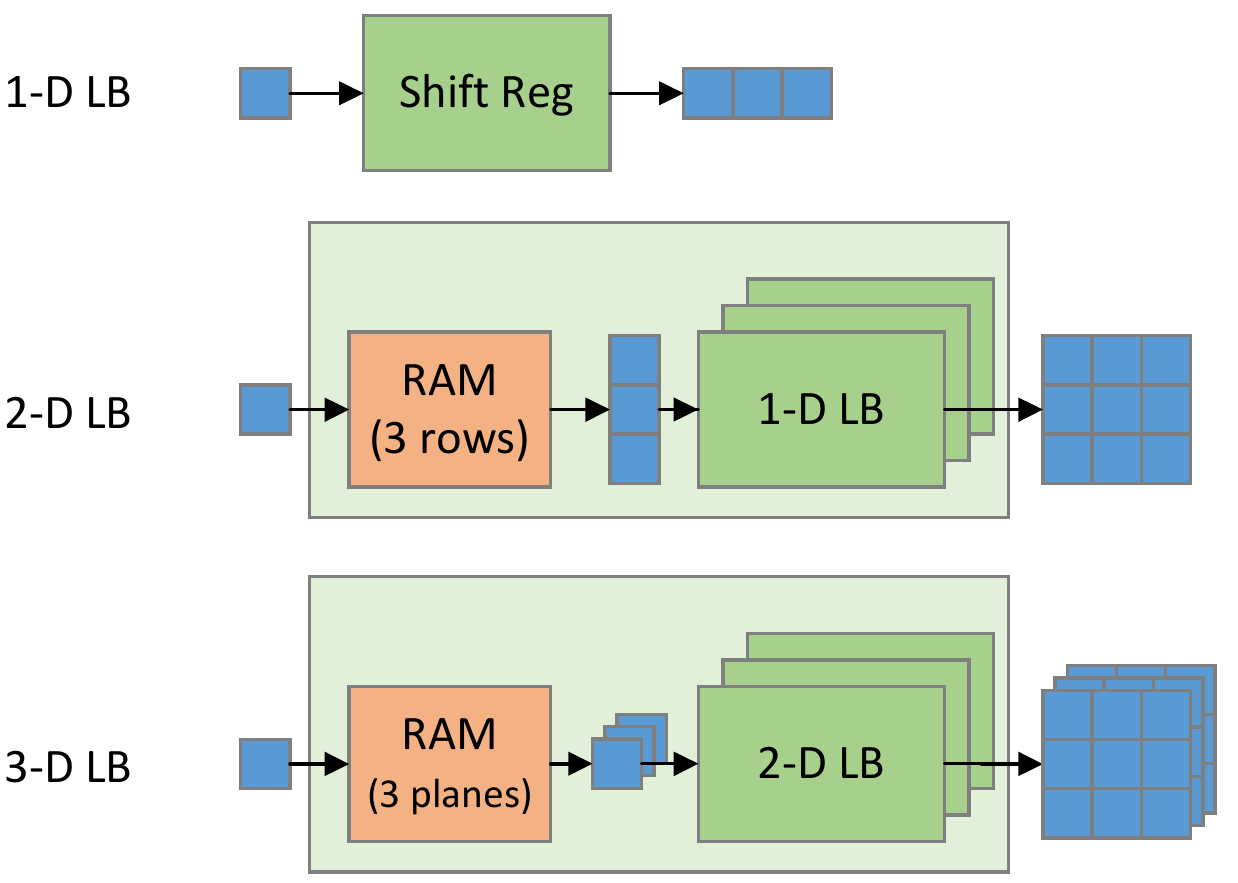}
    \caption{Multi-dimensional line buffer.
        1-D LB uses shift register.
        2-D LB uses RAM to buffer rows of pixels, pushes a column to 1-D LBs and outputs 2-D stencils.
        3-D LB instantiates RAMs and 2-D LBs similarly.
        }
    \label{fig:lb-impl}
    \vspace{-0pt}
\end{figure}

After some common optimizations,
the final IR of the pipeline is passed to two different code generator back-ends, HLS and LLVM.

The HLS code generator translates the hardware accelerator portions of IR into HLS-synthesizable C code,
and the rest of the IR is translated into a C++ testbench wrapper.
The code generator inserts HLS directives (pragmas) automatically to assist the HLS compiler in applying loop pipelining and array partitioning.
To simplify the code generator, we built an HLS-synthesizable 
C++ template library implementing an abstract line buffer interface:

\begin{lstlisting}[language=C++, morekeywords={stream}, frame=none, numbers=none]
template<int IMG_SIZE_0, int IN_SIZE_0, int OUT_SIZE_0, ...>
void linebuffer(stream<IN_SIZE_0, ...> &in,
                stream<OUT_SIZE_0, ...> &out);
\end{lstlisting}
\noindent
We design the multi-dimensional (up to 4D) line buffer templates hierarchically, as shown in \autoref{fig:lb-impl}.

Enabling more design optimizations, the compiler can also statically evaluate constant functions (e.g. lookup tables), and generate the code that later synthesizes to ROMs.

Generating machine code for the CPU portion of the software program is left to the LLVM compiler infrastructure.
We largely use the existing Halide ARM back-end, which 
includes a highly optimized ARM NEON SIMD vectorizer,
thread pool-based parallel runtime, etc.
The final IR describes a complete pipeline with both CPU and accelerator,
from which the code generator emits 
platform-specific device driver calls to access the hardware when it visits the boundary of the hardware pipeline during IR traversal.
Moreover, the tool recognizes all the data buffers accessed by 
any hardware pipeline, and thus can emit special allocation routines for these buffers,
inserting data transfers where required
by the platform setup.
\section{Platform Development}
\label{sec:platform}

\IGNORE{
\begin{figure}
    \centering
    \includegraphics[width=.7\columnwidth]{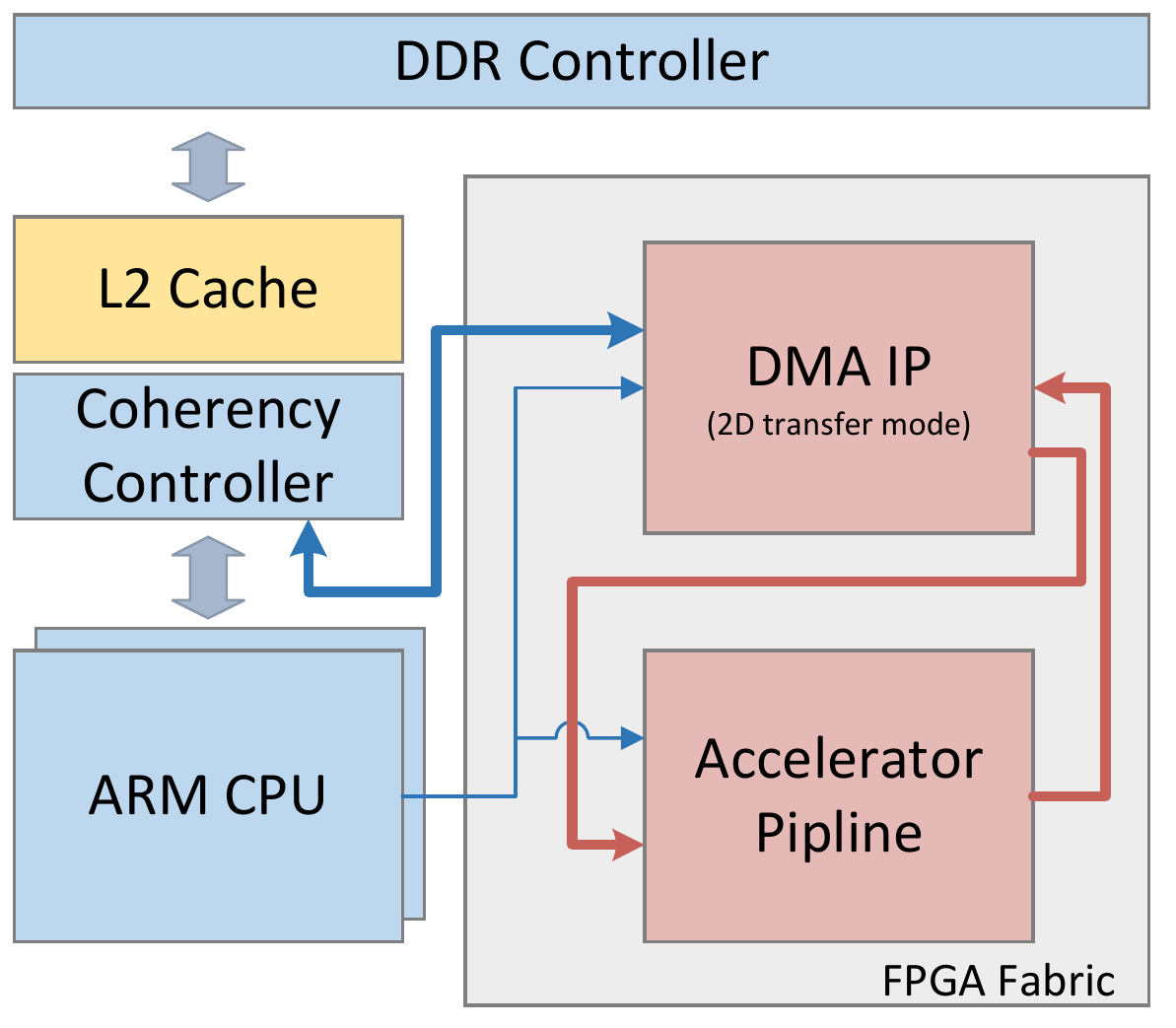}
    \caption{Zynq configuration. Accelerators and DMAs are programmed on FPGA. The DMAs are cache coherent and share a 512KB L2 with two ARM cores. Blue and red arrows represent AXI4 and AXI4-Stream buses, respectively.}
    \label{fig:zynq}
\end{figure}
}

We implemented our system on a Xilinx Zynq-7000 SoC platform~\cite{xilinx7000datasheet}, which
contains a pair of ARM Cortex A9 cores and FPGA fabric.
We run Linux on the ARMs, giving us the many conveniences of an operating system (e.g. a file system, networking, and core utilities), and realistically modeling the challenges of integrating an accelerator into a larger heterogeneous system.
The generated Halide program appears to the user as a single C ABI function, 
callable from ARM CPU userland;
all interaction with the FPGA fabric is automatically managed within that function.

We use AXI DMA from Xilinx~\cite{xilinx2015dma7v1} to connect the streaming interface of the accelerators to the CPU cache hierarchy through an accelerator coherency port (ACP).
The DMA engines use 2D transfer mode to access data row-by-row with a
fixed offset between the start of each row.
By adjusting the DMA configuration,
we can easily send a sub-image to the accelerator without having to copy or move any data.

We built a parametrized Linux kernel module to drive the DMA engines and the accelerators.
The driver provides a simple interface used by the generated Halide CPU code, including an asynchronous launch for the hardware pipeline and a synchronization barrier to wait for the hardware to complete.


We overlap the execution of CPU cores and accelerators using Linux threads.
After the output image has been tiled into smaller blocks,
the processing pipeline of the small block is wrapped by a loop iterating over the tiles.
A user can schedule the loop to run in parallel 
using Halide's \code{parallel} primitive,
as shown in line~16 of \autoref{fig:unsharp-code}.
After this, a typical software program looks like:

\begin{lstlisting}[morekeywords={parallel, retrieve, release, launch_accelerator, wait_sync}, 
% frame=none,
% linewidth=.7\columnwidth,
xleftmargin=.16\columnwidth, xrightmargin=.16\columnwidth,
frame=single,
numbers=none]
parallel foreach tile_index:
  retrieve pinned_buffers
  compute values in pinned_buffers...
  task_id = launch_accelerator(pinned_buffers)
  wait_sync(task_id)
  consume values in pinned_buffers...
  release pinned_buffer
\end{lstlisting}

\noindent
During execution, a thread pool is created for launching the workers that run the loop body.
Some threads use CPU cores to compute values in the buffer, launch the accelerator task,
and then quickly get blocked in the \code{wait\_sync} calls.
While these threads sleep and the accelerator is running, other active threads can use the CPU cores.
To avoid cache thrashing by too many contexts running concurrently,
the number of active threads is limited to three.
\section{Evaluation}
\label{sec:evaluation}

We now describe our experimental setup,
followed by evaluation on four separate fronts:
1) {\em hardware generation}, comparing our generated hardware versus optimized kernels from an HLS library and generated hardware from the HIPAcc compiler;
2) impact of the {\em loop perfection op\-ti\-mi\-za\-tion};
3) {\em heteroge\-ne\-ous system performance}, where we describe our efforts to optimize 
the hardware and software running on the target platform;
and 4) 
{\em efficient programmability},
where we show that we can generate competitive hardware and software for a variety of Halide applications.

\subsection{Experimental Setup}

\textbf{Compiler:} Our compiler is based on open source Halide~\cite{halide-homepage} built using GCC\,4.8 and LLVM\,3.6.
The evaluated implementations for the Zynq platform,
ARM CPU and CUDA GPU are all generated by our compiler,
although the CPU and GPU code is the same as
produced by the original Halide compiler.
We also compared the open source HIPAcc-Vivado compiler~\cite{hipaccvivado-git}.

\textbf{Applications:} \autoref{table:applications} lists the applications we use in this paper, including
{\em Gaussian}, a basic stencil kernel;
{\em Harris corner detector}, a pipeline of stencil kernels;
{\em unsharp mask}, an example of a DAG of kernels;
{\em stereo}, a compute intensive algorithm calculating stereo correspondence using block matching;
{\em bilateral grid}, a fast bilateral filtering algorithm (an edge-preserving smoothing filter) containing rich patterns of histogram, downsampling and data gathering~\cite{chen2007real, paris2009fast}; and
a simple {\em camera} pipeline from Frankencamera~\cite{Adams:2010:TFA}.
The last two applications are adapted from Halide's open source repository.

\begin{table}
  \small
  \centering
  \begin{tabular}{r l}
    \textbf{Application} & \textbf{Description}\\
    \midrule
    gaussian & 9-point 2-D Gaussian filter \\
    harris & Harris corner detector \\
    unsharp & unsharp masking filter \\
    stereo & stereo depths using block matching \\
    bilateral grid & fast bilateral filtering algorithm \\
    camera & camera pipeline from Frankencamera \\
    \bottomrule
  \end{tabular}
  \caption{Applications used in this paper.}
    \vspace{-0pt}
  \label{table:applications}
\end{table}

\textbf{Platforms:} We use a Xilinx ZC702 evaluation board holding a Zynq XC7Z020 SoC
as the target heterogeneous platform,
running a Linux\,4.0 kernel built from Xilinx Open Source Linux\,2015.4 ~\cite{xilinx-linux}
along with our custom built device driver module.
We use Xilinx Vivado Design Suite 2015.4 for synthesizing HLS, RTL simulation and generating an FPGA bitstream.

For comparison to a CPU+GPU platform, we use an NVIDIA Jetson TK1 board
with JetPack\,2.0~\cite{nvidia-jetpack}.
The Tegra K1 SoC is fabricated in the same 28nm technology as the Zynq SoC.
We evaluate both ARM and CUDA implementations on TK1 for each application using the same Halide algorithm but different schedules, optimized by Halide experts.
For {\em stereo,}
where Halide-generated CUDA performs poorly,
we compared to the Tegra-optimized OpenCV CUDA kernel shipped by NVIDIA with JetPack.

\textbf{Measuring power and performance:} 
We pull statistics from Texas Instruments UCD9248 power controllers on the ZC702,
which reports the power of each subsystem
including FPGA, DRAM, and CPU.
For the TK1 board, we measure the current on the 12V DC supply.
To derive the energy efficiency for TK1,
we subtract the board idle power (about 2 Watts) to exclude the board's uncore components,
which should give the TK1 an advantage as it also excludes SoC and DRAM static power.
We use \code{gettimeofday} to measure software program execution time and report the average over 20 runs.
For the CUDA target, we exclude the data transfer time between CPU and GPU (again giving the TK1 an advantage).

\subsection{Hardware Generation}
\label{sec:hw_gen_result}

\begin{figure}
    \centering
    \includegraphics[width=1.0\columnwidth]{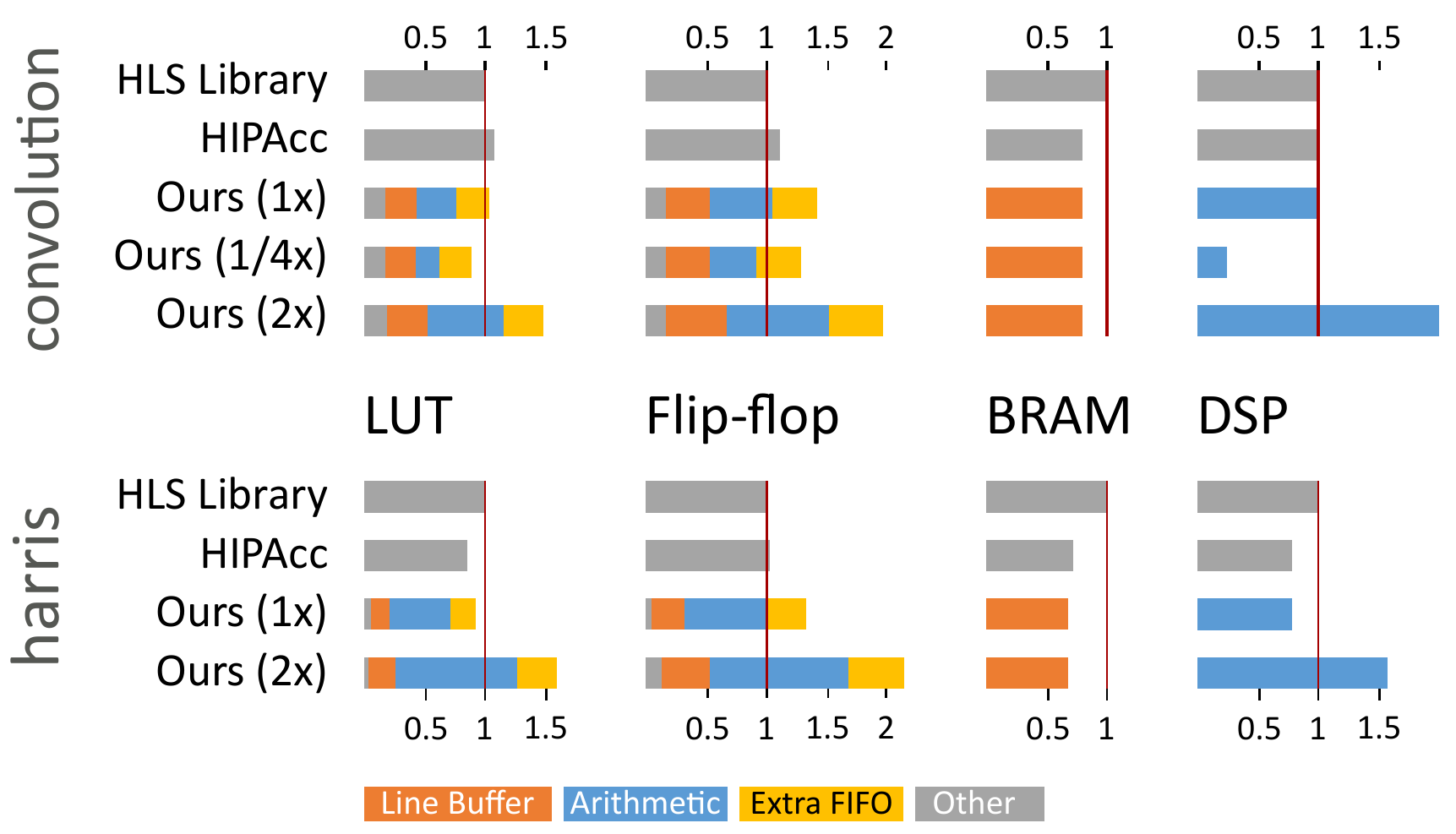}
    \caption{%
      Resource utilization comparison of \textit{conv2d} 
      (top)
      and \textit{harris} 
      (bottom)
      in our system at various pixel rates
      vs.\ HLS library and HIPAcc implementations.
      Our system's designs can target multiple different throughput,
      while HIPAcc and library kernels are only optimized at a 1 pixel/cycle rate. 
    }
    \label{fig:hardware_gen}
\end{figure}

From very compact high-level code, using a 5$\times$5 2-D convolution (\textit{conv2d}) and Harris as test cases, we find that our system can generate hardware quality similar to that of the Xilinx HLS video library~\cite{xilinx-hls} and HIPAcc-Vivado system.

\autoref{fig:hardware_gen} shows the FPGA resource utilization of our generated designs vs.~the library designs for \textit{conv2d} and \textit{harris} on a 1080p image.
We generated designs at different pixel rates using the same Halide algorithm
but different unrolling schedules, while the library and HIPAcc%
\footnote{
  The recent HIPAcc Altera-OpenCL backend~\cite{ozkan2016fpga} can generate >1 pixel/cycle pipelines through kernel vectorization, but cannot generate <1 pixel/cycle designs. We did not evaluate this system as it is not open-source.
}
only provide designs at a single pixel per cycle rate.
All the designs achieve similar peak frequencies, around 180MHz for \textit{conv2d} and 150MHz for \textit{harris}.
\textit{Harris} designs run slower because they are bounded by a floating-point-to-integer conversion.

Our line buffer components use less BRAM for two reasons.
First, the line buffer instance from the library is not optimal in terms of storage usage:
in 5$\times$5 convolution, the library design buffers five rows whereas the minimum required is four,
so both our and HIPAcc's designs start with 20\% less BRAM.
Second, the library and HIPAcc both instantiate per-kernel line buffers for each input stencil,
while we place a line buffer for each \emph{output} stencil stream,
which can be shared among kernels consuming the same stream.
In \textit{harris}, our design instantiates fewer line buffer instances thanks to this buffer sharing, for an extra 6\% BRAM savings as compared to HIPAcc.

In \textit{harris}, our design and HIPAcc use fewer DSPs and LUTs
because the code generation (meta-pro\-gram\-ming) approach is more flexible and does better constant propagation and simplification as compared to the C++ template solution used by the library.
%
Our unit-rate \textit{conv2d} design has fewer LUTs because the library solution creates four separate streams for the RGBA channels, whereas we package RGBA as a single structure and pass it through a wider stream along the pipeline,
which simplifies the control logic for managing data streams.

Multi-rate designs need higher compute throughput and buffer bandwidth.
Fortunately, the BRAM banks of the line buffers provide more than enough bandwidth for 1080p images.
Therefore, the multi-rate designs use more LUTs and DSPs for the
arithmetic datapath, while the line buffer resources do not change significantly.

One source of overhead in our design comes from unnecessary FIFOs inserted between pipeline stages.
We use \code{hls::stream} objects to connect different stages in the generated HLS code,
and the current HLS compiler creates a hardware FIFO for each stream object.
However, in our design, stages can be directly connected through a handshake interface (e.g. AXI4-stream)
because the latencies in the pipeline are already balanced.
The extra FIFO adds 30\% FF and 20\% LUT overheads.
Future optimizations in the HLS compilation could eliminate these unnecessary FIFOs.

\begin{table}
  \small
  \centering
  \begin{tabular}{ccccc}
    \toprule
&	\bf HLS Lib	& \bf HIPAcc & \bf Halide & \bf Generated HLS\\
    \midrule
\bf Conv2D &	209 &	5+5 &	2+2 & 885 \\
\bf Harris &	520 &	47+26 &	23+11 & 619 \\
    \bottomrule
  \end{tabular}
  \caption{
    Lines of code (LoC) for \textit{conv2d} and \textit{harris} in HLS library, HIPAcc, Halide and generated HLS code from Halide. HLS Lib excludes basic data structure code; HIPAcc counts are DSL parameter+algorithm~\cite{membarth2016hipa}; Halide counts are algorithm+schedule; Generated HLS counts exclude 900 LoC in the line buffer template library.
  }
  \label{table:code_length}
\end{table}

\autoref{table:code_length} summarizes the code length of \textit{conv2d} and \textit{harris} using different systems. Moving from less to more domain specific, Halide and HIPAcc DSLs are orders of magnitude more compact than both HLS C library and our generated HLS code. Because Halide uses functional representation, its application code is 2x shorter than HIPAcc's. 




\subsection{Impact of Loop Perfection}
\label{sec:loop_perfection_result}

\begin{table}
  \small
  \centering
  \begin{tabular}{lccc}
    \toprule
    ~       &	\bf Sched. rate	& \bf Meas. rate    & \bf Resource \\
\bf Application &	(pix/cyc)	&  (pix/cyc) & (LUT+FF) \\
    \midrule
\bf histogram     &	0.016 &	0.015 &	1.1\%+0.6\% \\
\bf histogram+opt &	0.016 &	0.016 &	2.0\%+1.0\% \\
\bf stereo        & 0.25	 &	0.062 &	55\%+25\%  \\
\bf stereo+opt    & 0.25	 &	0.25 &	55\%+32\%  \\
    \bottomrule
  \end{tabular}
  \caption{Scheduled and measured throughput and resource utilization of \textit{histogram} and \textit{stereo} with and without the loop perfection optimization.}
  \label{table:loop_perfection_result}
\end{table}

As discussed in \autoref{sec:loop_perfection},
the loop perfection optimization helps fully pipeline the 
stages that have sequential loops (running at <1 pixel/cycle rate).
Such pipeline stages are important in applications that require data-dependent reductions, e.g. the \textrm{histogram} stage in \textit{bilateral grid}, or are intentionally scheduled at slower rate due FPGA resource constraints, e.g. \textit{stereo}.

\autoref{table:loop_perfection_result} lists the scheduled and measured throughput and resource utilization of \textit{histogram} and \textit{stereo} with and without the loop perfection optimization.
The optimization improves the performance by 7\% and 300\% with low resource overhead for \textit{histogram} and \textit{stereo}, respectively.
\textit{stereo} gains more because the innermost loop is short (4 iterations) and the overhead of entering and exiting the short loop is relatively large if not pipelined with the outer loops.

\subsection{Heterogeneous System Performance}
\label{sec:blocking_result}

\begin{figure}
    \centering
    \includegraphics[width=.95\columnwidth]{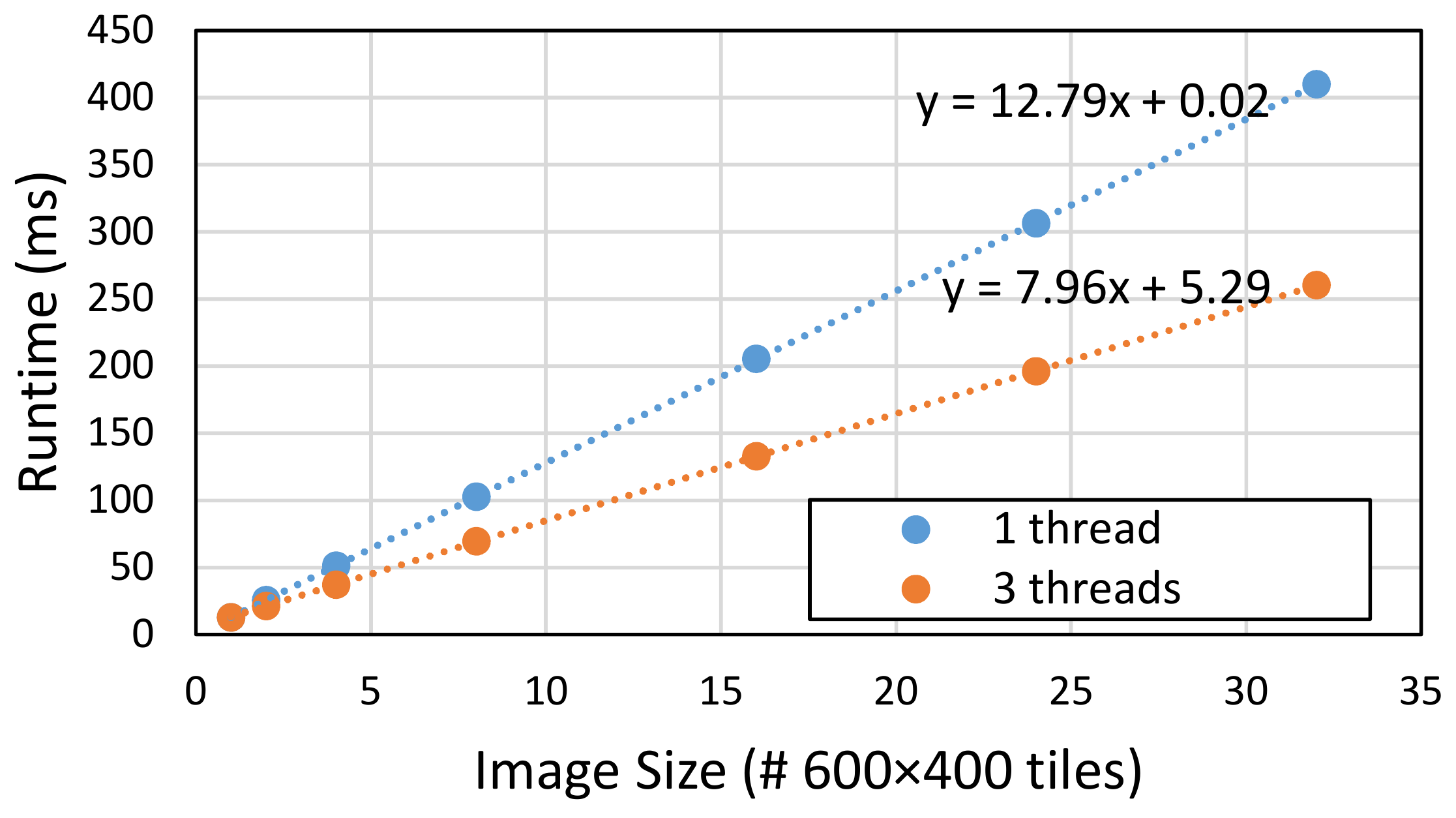}
    \caption{Runtime of \textit{stereo} with different sizes of input image.
    All runs use a hardware accelerator that processes a 600$\times$400 image tile
    in 7.93ms. Our multithreading technique helps overlap CPU and accelerator execution time, 
    so that per-tile processing cost plus CPU overhead becomes just 7.96ms instead of 12.76ms.
    }
    \label{fig:stereo_multithread}
\end{figure}



System performance can be greatly improved if the CPU and the accelerator run concurrently
while processing a heterogeneous pipeline.
\autoref{fig:stereo_multithread} illustrates the effectiveness of our multi-threading technique
from \autoref{sec:platform},
overlapping CPU and accelerator workloads for stereo.
The figure plots the program runtime using the same accelerator to process different size images with and without multiple threads.
For each launch, the accelerator processes a 600$\times$400 image tile in 7.63ms.
The software program breaks the image into such tiles,
prepares an input image tile for the accelerator including pad\-ding the boundary of the original image,
launches and waits for the accelerator, and then repeats for the next tile.

When running the program sequentially with one thread,
each tile takes 12.8ms, indicating a CPU workload of about 5ms/tile.
However, using 3 threads to process 3 tiles concurrently, the per-tile processing cost
reduces to 7.96ms, as most of the CPU workload now overlaps with accelerator execution.
With this overlapping, the system is bounded by its slowest part which in this case is the accelerator portion of the workload.  When the CPU workload dominates, we may not see such nice behavior, as we find out below.

On our target platform, the accelerators are cache coherent and
share the L2 cache with the CPU cores.
If intermediate buffers passing data between CPU kernels and accelerators fit in the L2,
the number of DRAM accesses will decrease, and the memory access latency will improve as well.
To this end, blocking the image via the \code{tile} primitive effectively reduces the size of the intermediate buffers.

\autoref{table:blocking} summarizes the performance and energy cost of \textit{gaussian} compiled for different block sizes and pixel rates.
We show the accelerator execution time for both Verilog simulation and
real time as measured when running the accelerator repeatedly on the Zynq.
Memory accesses start to miss in the L2 when the block size exceeds 48KB, 
causing the DRAM dynamic energy to increase from zero to around 3.3nJ/pixel (i.e. 34pJ/bit-access).\footnotemark

\footnotetext{
Computing an RGB pixel in gaussian requires 6 bytes of memory access.
If the memory access misses in the L2, we assume it causes two DRAM accesses.
}

\begin{table}
  \small
  \centering
  \begin{tabular}{cccccc}
    \toprule
    ~           & \bf Read &	\bf Block & \multicolumn{2}{c}{\bf ---HW time---}   &	\bf DRAM \\
    \bf Rate & \bf BW & \bf size &	\bf sim. &	\hspace{-10pt} \bf measured  &  \bf $E_{dyn}$\\
    (px/cy) & (MB/s) & (KB) & ($\mathrm{\mu s}$)	& ($\mathrm{\mu s}$)	& (nJ/px) \\
    \midrule
    \multirow{3}{*}{1} & \multirow{3}{*}{300} & ~48 &	~193 &	~199  &	0.0\\
	& & 192 &	~710 &	~721 &	2.5\\
	& & 768 &	2730 &	2764 &	3.3\\
    \midrule
    \multirow{3}{*}{2} & \multirow{3}{*}{600} & ~48 & ~~99 &	~111  &	0.0 \\
	& & 192 &	~360	& ~510 &	2.3\\
	& & 768 &	1378 &	2444 &	3.3\\
    \bottomrule
  \end{tabular}
  \caption{Performance and energy cost measurements of the \textit{gaussian} accelerators for different block sizes and pixel rates running at 100MHz.
    Larger blocks cause L2 cache misses to DRAM, increasing DRAM dynamic energy.
  \label{table:blocking}}
\end{table}

When we compare simulated to measured accelerator performance ({\em sim} vs.\ {\em measured} in \autoref{table:blocking}), we see that, for
low rate configurations, the measured runtime does not degrade 
when large block sizes cause L2 misses,
since DMA buffers can hide the latency.
However, for high rate configurations, although the accelerators are 
twice as fast than the unit-rate designs in simulation, peak speed is only achieved for the 48KB block configuration (i.e., when memory accesses all hit in the L2). Because of cache misses, the actual runtime is 77\% longer than the simulated accelerator runtime for the much larger 768KB block configuration.

However, breaking images into many small blocks introduces two problems of its own:
First, more overlapping block boundaries causes more re-computation;
and second, the host CPU has to schedule more accelerator launches through the device driver interface.
On the current platform, we observe scheduling overhead around 
100$\sim$200 microseconds
per launch in the device driver, mostly caused by context switches and synchronizations
between the user thread and kernel background threads responsible
for managing accelerator launch and completion queues.
As a result of these overheads, for most of the evaluated applications, fine-grained blocking to fit in the L2 is not an efficient strategy overall.
Instead, we chose to block the image in larger sizes (e.g. 480$\times$640) for our full system evaluation.

With further engineering of our existing system, it would be possible to increase memory bandwidth for streaming large blocks.
However, if we were to design a new platform, we would choose a faster CPU core (the Zynq's 667MHz A9 was slow even at its introduction in 2012),
a larger shared L2 cache,
and a hardware launch engine that pulls accelerator tasks directly from a memory buffer without CPU intervention,
as opposed to having a background thread pushing tasks to the accelerator.

\subsection{Programmability and Efficiency}
\label{sec:final_result}

\begin{table}
  \small
  \centering
  \begin{tabular}{rccc}
    \toprule
    & \bf Rate & \bf Read BW & \bf Resource \\
    \bf Application & (pix/cyc) & (MB/s) & (\%) \\
    \midrule
    \bf gaussian & 2 & 200 & 9/5/3/33 \\
    \bf harris & 2 & 284 & 22/19/7/30 \\
    \bf unsharp & 1 & 375 & 7/5/7/31 \\
    \bf stereo & 0.25 & 62 & 55/32/5/0 \\
    \bf bilateral & 1 & 182 & 23/20/14/13\\
    \bf camera & 2 & 568 & 9/6/5/5 \\
    \bf camera+unsharp & 1 & 250 & 14/10/11/34\\
    \bottomrule
  \end{tabular}
  \caption{Specifications of generated accelerators for the evaluated applications.  The resource utilization is reported as percentage of LUT, FF, BRAM, and DSP used by the given application on the Zynq XC7Z020. }
  \label{table:accelerators}
\end{table}

\begin{figure}
    \centering
    \includegraphics[width=1\columnwidth]{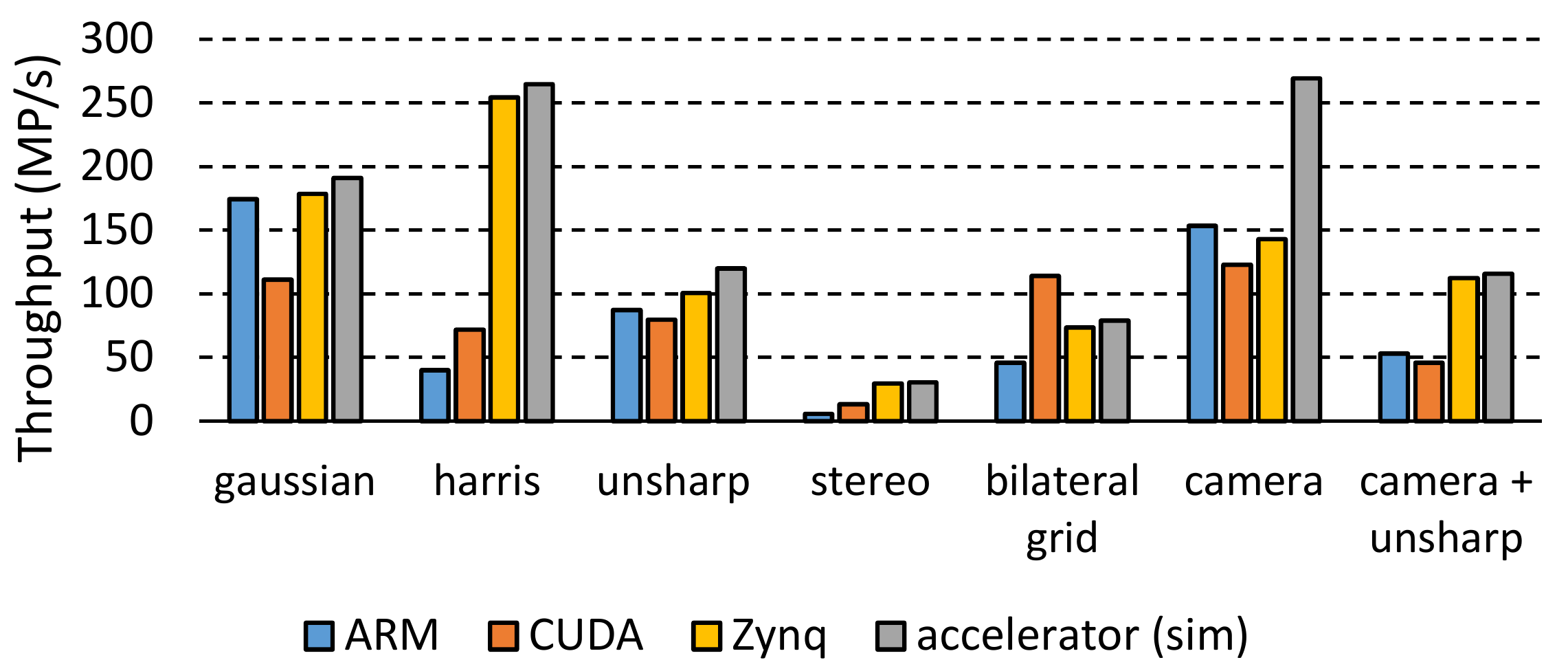}
    \includegraphics[width=1\columnwidth]{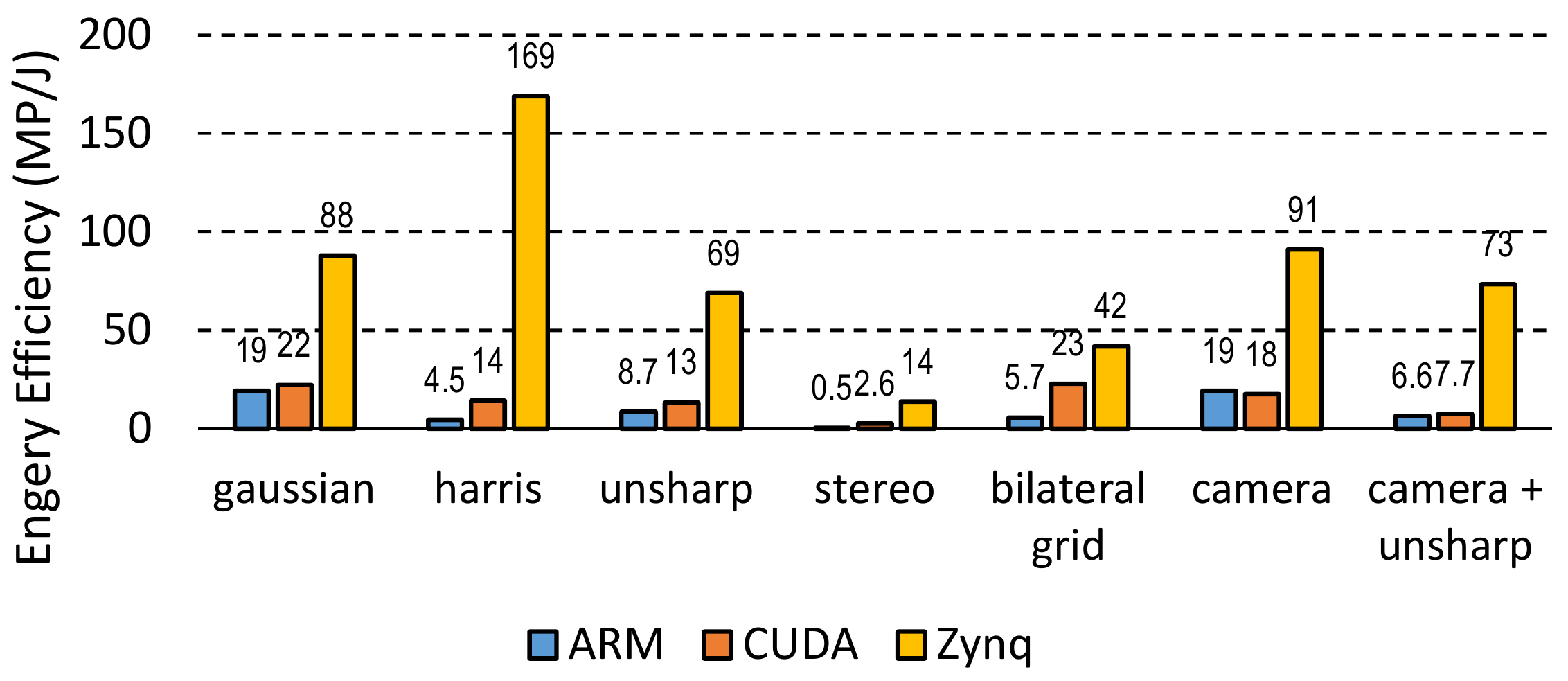}
    \caption{
      Throughput and energy efficiency comparison of Zynq platform versus the four ARM cores and CUDA GPU on TK1. The accelerator RTL simulation throughput represents the theoretical ideal for the synthesized accelerator in isolation, if it were never bottlenecked on other parts of the system (namely input and output bandwidth and latency). Additional system optimization, and improved CPU performance in future SoCs, could push realized Zynq performance closer to this level.
    }
    \label{fig:throughput}
\end{figure}

We implemented six individual applications and 
an application that combined the camera pipeline with an unsharp mask, all in Halide,
and generated the hardware and software for Zynq using our compiler.

\autoref{table:accelerators} lists the specifications of generated accelerators for these applications.
For most of the applications, the CPU just does tiling, i.e.~calculating the coordinates of each image tile and scheduling the accelerator for processing a tile.
In \textit{stereo} the CPU computes padding 
using a repeat edge condition, and in \textit{bilateral grid} it shuffles data into 8$\times$8 grid order.
We run 8-megapixel images through each application, but the BRAM usage for internal buffering in each accelerator is kept low thanks to the image tiling.

\autoref{fig:throughput} shows the throughput and energy efficiency of the Zynq implementation versus TK1's ARM and GPU cores.
On average, applications on Zynq achieve 2.6$\times$ and 1.9$\times$ higher throughput, and 14.2$\times$ and 6.1$\times$ higher energy efficiency compared to the CPUs and GPU on the TK1, respectively.
\textit{Harris} achieves the most energy reduction of 38$\times$ and 12$\times$,
as well as
the highest throughput speedups of 6$\times$ and 3.5$\times$, compared to the TK1 CPUs and GPU.
The energy efficiency is achieved by high locality and data reuse exploited in the line-buffered accelerator pipeline,
and the greatly reduced memory requests as compared to programmable cores.
Low-precision fixed-point arithmetic is also very efficient in LUTs and DSPs on FPGA fabric.

All Zynq-based applications except \textit{camera}
achieve the peak throughput of the accelerator in Verilog simulation.
The \textit{camera} accelerator requires much higher read bandwidth,
and thus suffers the most from bandwidth problems caused by
reading data that misses in L2 cache (see \autoref{sec:blocking_result}).

Once the application fits in the FPGA fabric, the throughput of Zynq implementations is generally bound
by memory bandwidth and clock frequency.
Therefore, speedups compared to the ARM CPUs or GPU on TK1 are proportional to the number of operations accelerated on the FPGA
(approximately proportional to the LUTs and the DSPs used).
For this reason, \textit{harris}, \textit{stereo}, and \textit{camera+unsharp} get the most speedup.
In \textit{bilateral grid}, the accelerator also uses a lot of LUTs, but most of them implement control logic
and multiplexers for building histograms and data-gathering for interpolation, which do relatively little real computation.
Moreover, parallelism in the kernels of the \textit{bilateral grid} is limited by data dependencies,
making it more favorable to execute on high frequency and high memory bandwidth processors (i.e., the Tegra CPUs and GPU).

An important takeaway is that the acceleration using the FPGA becomes more effective as the image processing pipeline grows deeper,
which matches the trend of new applications from computational photography and computer vision.
For example, on CPU or GPU, the execution time of the \textit{camera+unsharp} combination
is the sum of the execution times of the two individual applications.
However, pipelining two applications onto the FPGA fabric simultaneously doesn't increase the required memory bandwidth or slow the clock frequency.
Therefore, the throughput for any composition of pipelines which fit on the FPGA is bounded by the slowest one (in this case, \textit{unsharp}, which is around 110 megapixels/second).

\subsection{Extensibility}

Our dataflow IR provides an untimed, bit accurate specification of image processing pipelines
with explicit coarse grain pipeline parallelism information, so the system can be easily extended
to target other HLS compilation tools simply by providing new code generators.
To demonstrate this extensibility, we ported the system to Catapult HLS~\cite{catapult}, a popular commercial tool for ASIC technology, 
by changing just around 1000 lines of code in the existing 
code generator module.
We evaluated each ~\autoref{table:applications} application in a 14nm ASIC technology using the new backend.
The hardened pipelines in ASIC get approximately 12$\times$ higher throughput and 8$\sim$20$\times$ better energy efficiency than those programmed on the Zynq FPGA (fabricated in 28nm).

\section{Conclusion}
\label{sec:conclusion}

The Halide image processing DSL provides an ability to quickly create and optimize new image processing applications.  The availability of complex SoC chips with large FPGA fabrics provides a potential platform for exploring new heterogeneous architectures for these applications, but  implementing the hardware accelerator and the interface software is a huge barrier for many designers. 
We extended Halide to remove this barrier, allowing it to generate both the design of the accelerator and the software that communicates with the hardware.  Our results demonstrate significant gains in both performance and energy efficiency, which  increase as the imaging computation becomes more complex.  We will open source our system to share with the community for others to use and build on.\footnote{The source is available at {\hypersetup{urlcolor=blue} \url{https://github.com/jingpu/Halide-HLS}.}}

We also plan to extend our system in several ways.  First, we will incorporate more automatic optimization, to help the designer find optimal low-level buffering and blocking. Next, given the generality of the underlying computational model, we want to extend this system to generate ``code" for different underlying image processing engines 
including custom hardware and the specialized SIMD and coarse-grain-reconfigurable-array
architectures optimized for image processing appearing in new SoCs. We hope this system will help the community develop and use efficient programmable ISPs.

\clearpage


\bibliographystyle{plain}
\bibliography{ref}

\begin{thebibliography}{10}

\bibitem{Adams:2010:TFA}
Andrew Adams, Eino-Ville Talvala, Sung~Hee Park, David~E. Jacobs, Boris Ajdin,
  Natasha Gelfand, Jennifer Dolson, Daniel Vaquero, Jongmin Baek, Marius Tico,
  Hendrik P.~A. Lensch, Wojciech Matusik, Kari Pulli, Mark Horowitz, and Marc
  Levoy.
\newblock The {Frankencamera}: An experimental platform for computational
  photography.
\newblock {\em ACM Transactions on Graphics}, 29(4):29:1--29:12, July 2010.

\bibitem{auerbach2012lime}
Joshua Auerbach, David~F. Bacon, Ioana Burcea, Perry Cheng, Stephen~J. Fink,
  Rodric Rabbah, and Sunil Shukla.
\newblock A compiler and runtime for heterogeneous computing.
\newblock In {\em Proceedings of the 49th Annual Design Automation Conference},
  DAC '12, pages 271--276, New York, NY, USA, 2012. ACM.

\bibitem{auerbach2010lime}
Joshua Auerbach, David~F Bacon, Perry Cheng, and Rodric Rabbah.
\newblock Lime: a {Java}-compatible and synthesizable language for
  heterogeneous architectures.
\newblock In {\em ACM Sigplan Notices}, volume~45, pages 89--108. ACM, 2010.

\bibitem{john2015stencil}
John~S. Brunhaver.
\newblock {\em Design and Optimization of a Stencil Engine}.
\newblock PhD thesis, Stanford University, 2015.

\bibitem{chen2007real}
Jiawen Chen, Sylvain Paris, and Fr{\'e}do Durand.
\newblock Real-time edge-aware image processing with the bilateral grid.
\newblock In {\em ACM Transactions on Graphics (TOG)}, volume~26, page 103.
  ACM, 2007.

\bibitem{HLS4FPGA}
J.~Cong, B.~Liu, S.~Neuendorffer, J.~Noguera, K.~Vissers, and Z.~Zhang.
\newblock High-level synthesis for {FPGAs}: From prototyping to deployment.
\newblock {\em IEEE Transactions on Computer-Aided Design of Integrated
  Circuits and Systems}, 30(4):473--491, April 2011.

\bibitem{czajkowski2012opencl}
Tomasz~S Czajkowski, Utku Aydonat, Dmitry Denisenko, John Freeman, Michael
  Kinsner, David Neto, Jason Wong, Peter Yiannacouras, and Deshanand~P Singh.
\newblock From {OpenCL} to high-performance hardware on {FPGA}s.
\newblock In {\em Field Programmable Logic and Applications (FPL), 2012 22nd
  International Conference on}, pages 531--534. IEEE, 2012.

\bibitem{george2014hardware}
Nivia George, HyoukJoong Lee, David Novo, Tiark Rompf, Kevin~J Brown, Arvind~K
  Sujeeth, Martin Odersky, Kunle Olukotun, and Paolo Ienne.
\newblock Hardware system synthesis from domain-specific languages.
\newblock In {\em Field Programmable Logic and Applications (FPL), 2014 24th
  International Conference on}, pages 1--8. IEEE, 2014.

\bibitem{govindaraju2011dynamically}
Venkatraman Govindaraju, Chen-Han Ho, and Karthikeyan Sankaralingam.
\newblock Dynamically specialized datapaths for energy efficient computing.
\newblock In {\em High Performance Computer Architecture (HPCA), 2011 IEEE 17th
  International Symposium on}, pages 503--514. IEEE, 2011.

\bibitem{catapult}
Mentor Graphics.
\newblock Catapult high-level synthesis.
\newblock \url{https://www.mentor.com/hls-lp/catapult-high-level-synthesis/}.

\bibitem{halide-homepage}
Halide.
\newblock Halide, a language for image processing and computational
  photography.
\newblock \url{http://halide-lang.org/}.

\bibitem{hauck2010reconfigurable}
Scott Hauck and Andre DeHon.
\newblock {\em Reconfigurable computing: the theory and practice of FPGA-based
  computation}.
\newblock Morgan Kaufmann, 2010.

\bibitem{hegarty2014siggraph}
James Hegarty, John Brunhaver, Zachary DeVito, Jonathan Ragan-Kelley, Noy
  Cohen, Steven Bell, Artem Vasilyev, Mark Horowitz, and Pat Hanrahan.
\newblock Darkroom: Compiling high-level image processing code into hardware
  pipelines.
\newblock {\em ACM Trans. Graph.}, 33(4):144:1--11, July 2014.

\bibitem{martin2009hls}
Grant Martin and Gary Smith.
\newblock High-level synthesis: Past, present, and future.
\newblock {\em IEEE Design \& Test of Computers}, 26(4):18--25, 2009.

\bibitem{mei2003adres}
Bingfeng Mei, Serge Vernalde, Diederik Verkest, Hugo De~Man, and Rudy
  Lauwereins.
\newblock Adres: An architecture with tightly coupled {VLIW} processor and
  coarse-grained reconfigurable matrix.
\newblock In {\em Field Programmable Logic and Application}, pages 61--70.
  Springer, 2003.

\bibitem{hipaccvivado-git}
Richard Membarth and Oliver Reiche.
\newblock Fork of {HIPAcc} generating code for {Vivado HLS}.
\newblock \url{https://github.com/hipacc/hipacc-vivado}.

\bibitem{membarth2016hipa}
Richard Membarth, Oliver Reiche, Frank Hannig, J{\"u}rgen Teich, Mario
  K{\"o}rner, and Wieland Eckert.
\newblock {HIPAcc}: A domain-specific language and compiler for image
  processing.
\newblock {\em IEEE Transactions on Parallel and Distributed Systems},
  27(1):210--224, 2016.

\bibitem{milder2012spiral}
Peter Milder, Franz Franchetti, James~C Hoe, and Markus P{\"u}schel.
\newblock Computer generation of hardware for linear digital signal processing
  transforms.
\newblock {\em ACM Transactions on Design Automation of Electronic Systems
  (TODAES)}, 17(2):15, 2012.

\bibitem{nickolls2008cuda}
John Nickolls, Ian Buck, Michael Garland, and Kevin Skadron.
\newblock Scalable parallel programming with {CUDA}.
\newblock {\em Queue}, 6(2):40--53, 2008.

\bibitem{nvidia-jetpack}
NVIDIA.
\newblock Jetpack for {L4T}.
\newblock \url{https://developer.nvidia.com/embedded/jetpack}.

\bibitem{owaida2011synthesis}
Muhsen Owaida, Nikolaos Bellas, Konstantis Daloukas, and Christos~D
  Antonopoulos.
\newblock Synthesis of platform architectures from {OpenCL} programs.
\newblock In {\em Field-Programmable Custom Computing Machines (FCCM), 2011
  IEEE 19th Annual International Symposium on}, pages 186--193. IEEE, 2011.

\bibitem{ozkan2016fpga}
M~Akif {\"O}zkan, Oliver Reiche, Frank Hannig, and J{\"u}rgen Teich.
\newblock Fpga-based accelerator design from a domain-specific language.
\newblock In {\em International Conference on Field Programmable Logic and
  Applications}, 2016.

\bibitem{papakonstantinou2009fcuda}
Alexandros Papakonstantinou, Karthik Gururaj, John~A Stratton, Deming Chen,
  Jason Cong, and Wen-Mei~W Hwu.
\newblock {FCUDA}: Enabling efficient compilation of {CUDA} kernels onto
  {FPGA}s.
\newblock In {\em Application Specific Processors, 2009. SASP'09. IEEE 7th
  Symposium on}, pages 35--42. IEEE, 2009.

\bibitem{paris2009fast}
Sylvain Paris and Fr{\'e}do Durand.
\newblock A fast approximation of the bilateral filter using a signal
  processing approach.
\newblock {\em International journal of computer vision}, 81(1):24--52, 2009.

\bibitem{prabhakar2015generating}
Raghu Prabhakar, David Koeplinger, Kevin Brown, HyoukJoong Lee, Christopher
  De~Sa, Christos Kozyrakis, and Kunle Olukotun.
\newblock Generating configurable hardware from parallel patterns.
\newblock {\em arXiv preprint arXiv:1511.06968}, 2015.

\bibitem{qualcomm800}
{Qualcomm Inc.}
\newblock Snapdragon 800 series mobile processors.
\newblock \url{https://www.qualcomm.com/products/snapdragon/processors/}
  \url{800-series}.

\bibitem{ragan2012decoupling}
Jonathan Ragan-Kelley, Andrew Adams, Sylvain Paris, Marc Levoy, Saman
  Amarasinghe, and Fr{\'e}do Durand.
\newblock Decoupling algorithms from schedules for easy optimization of image
  processing pipelines.
\newblock {\em ACM Transactions on Graphics (TOG)}, 31(4):32, 2012.

\bibitem{ragankelley2013pldi}
Jonathan Ragan-Kelley, Connelly Barnes, Andrew Adams, Sylvain Paris, Fr{\'e}do
  Durand, and Saman Amarasinghe.
\newblock Halide: A language and compiler for optimizing parallelism, locality,
  and recomputation in image processing pipelines.
\newblock In {\em Proceedings of the 34th ACM SIGPLAN Conference on Programming
  Language Design and Implementation}, PLDI '13, pages 519--530, New York, NY,
  USA, 2013. ACM.

\bibitem{hipaccvivado}
Oliver Reiche, Moritz Schmid, Frank Hannig, Richard Membarth, and J{\"u}rgen
  Teich.
\newblock Code generation from a domain-specific language for {C-based HLS} of
  hardware accelerators.
\newblock In {\em Hardware/Software Codesign and System Synthesis (CODES+
  ISSS), 2014 International Conference on}, pages 1--10. IEEE, 2014.

\bibitem{stone2010opencl}
John~E Stone, David Gohara, and Guochun Shi.
\newblock {OpenCL}: A parallel programming standard for heterogeneous computing
  systems.
\newblock {\em Computing in science \& engineering}, 12(1-3):66--73, 2010.

\bibitem{sujeeth2011optiml}
Arvind Sujeeth, HyoukJoong Lee, Kevin Brown, Tiark Rompf, Hassan Chafi, Michael
  Wu, Anand Atreya, Martin Odersky, and Kunle Olukotun.
\newblock {OptiML}: an implicitly parallel domain-specific language for machine
  learning.
\newblock In {\em Proceedings of the 28th International Conference on Machine
  Learning (ICML-11)}, pages 609--616, 2011.

\bibitem{Wolfe:1992:BIV:143095.143131}
Michael Wolfe.
\newblock Beyond induction variables.
\newblock In {\em Proceedings of the ACM SIGPLAN 1992 Conference on Programming
  Language Design and Implementation}, PLDI '92, pages 162--174, New York, NY,
  USA, 1992. ACM.

\bibitem{xilinx2015dma7v1}
Xilinx.
\newblock {AXI DMA v7.1 LogiCORE IP} product guide.
\newblock
  \url{http://www.xilinx.com/support/documentation/ip_documentation/axi}
  \url{_dma/v7_1/pg021_axi_dma.pdf}.

\bibitem{xilinx-hls}
Xilinx.
\newblock Vivado high-level synthesis.
\newblock
  \url{http://www.xilinx.com/products/design-tools/vivado/integration/esl-design.html}.

\bibitem{xilinx-linux}
Xilinx.
\newblock Xilinx wiki - open source {Linux}.
\newblock \url{http://www.wiki.xilinx.com/Open+Source+Linux}.

\bibitem{xilinx7000datasheet}
Xilinx.
\newblock Zynq-7000 all programmable {SoC} overview.
\newblock \url{http://www.xilinx.com/support/documentation/data_sheets/ds190-}
  \url{Zynq-7000-Overview.pdf}.

\bibitem{zhang2008autopilot}
Zhiru Zhang, Yiping Fan, Wei Jiang, Guoling Han, Changqi Yang, and Jason Cong.
\newblock Auto{P}ilot: A platform-based {ESL} synthesis system.
\newblock In {\em High-Level Synthesis}, pages 99--112. Springer, 2008.

\end{thebibliography}


\end{document}